\pdfoutput=1
\RequirePackage{ifpdf}
\ifpdf 
\documentclass[pdftex]{sigma}
\else
\documentclass{sigma}
\fi

\numberwithin{equation}{section}
\newtheorem{Theorem}{Theorem}[section]
\newtheorem{Corollary}[Theorem]{Corollary}
\newtheorem{Proposition}[Theorem]{Proposition}
{ \theoremstyle{definition}
\newtheorem{Remark}[Theorem]{Remark}}

\begin{document}

\newcommand{\arXivNumber}{1409.5444}

\allowdisplaybreaks

\renewcommand{\PaperNumber}{028}

\FirstPageHeading

\ShortArticleName{Darboux Transformations for $(2+1)$-Dimensional Extensions of the KP Hierarchy}

\ArticleName{Darboux Transformations for $\boldsymbol{(2+1)}$-Dimensional\\
Extensions of the KP Hierarchy}

\Author{Oleksandr CHVARTATSKYI~$^\dag$ and Yuriy SYDORENKO~$^\ddag$}

\AuthorNameForHeading{O.~Chvartatskyi and Yu.~Sydorenko}

\Address{$^\dag$~Mathematisches Institut, Georg-August-Universit\"at G\"ottingen, 37073 G\"ottingen, Germany}
\EmailD{\href{mailto:alex.chvartatskyy@gmail.com}{alex.chvartatskyy@gmail.com}}

\Address{$^\ddag$~Faculty of Mechanics and Mathematics, Ivan Franko National University of Lviv,\\
\phantom{$^\ddag$}~79000 Lviv, Ukraine}
\EmailD{\href{mailto:y_sydorenko@franko.lviv.ua}{y\_sydorenko@franko.lviv.ua}}

\ArticleDates{Received September 23, 2014, in f\/inal form March 27, 2015; Published online April 10, 2015}

\Abstract{New extensions of the KP and modif\/ied KP hierarchies with self-consistent sources are proposed.
The latter provide new generalizations of $(2+1)$-dimensional integrable equations, including the DS-III equation and
the~$N$-wave problem.
Furthermore, we recover a~system that contains two types of the KP equation with self-consistent sources as special
cases.
Darboux and binary Darboux transformations are applied to generate solutions of the proposed hierarchies.}

\Keywords{KP hierarchy; symmetry constraints; binary Darboux transformation; Davey--Stewartson equation; KP equation
with self-consistent sources}

\Classification{35Q51; 35Q53; 35Q55; 37K35}

\section{Introduction}

In the past years, a~lot of attention have been given to the study of Kadomtsev--Petviashvili hierarchy (KP hierarchy)
and its generalizations from both physical and mathematical points of
view~\cite{MAblowitz,LDA,DJM,NovikovZakharov,Ohta,SS3}.
KP equation with self-consistent sources and related~$k$-constrained KP ($k$-cKP) hierarchy also present an
interest~\cite{ANP,Chenga1,Chenga2,HelminckLeur,KSS,Krich,MV88,MV89,Orlov-2,Orlov-1,Orlov-3,SD}.
The latter hierarchy contains, in particular, nonlinear Schr\"odinger equation, Yajima--Oikawa equation, extension of
the Boussinesq equation and KdV equation with self-consistent sources.
A~modif\/ied~$k$-constrained KP ($k$-cmKP) hierarchy was proposed in~\cite{CY,KSO,OC}.
The~$k$-cKP hierarchy was extended to 2+1 dimensions ($(2+1)$-dimensional~$k$-cKP hierarchy) in~\cite{LZL1, MSS,6SSS}.

A powerful solution generating method for nonlinear integrable systems is based on the Darboux transformations (DT) and
the binary Darboux transformations (BDT)~\cite{Matveev}.
The latter transformations were also applied to~$k$-cKP hierarchy and its $(2+1)$-dimensional extensions
(see~\cite{Oevel93,Will97} and~\cite{LZL2, 6SSS} respectively).
More general $(2+1)$-dimensional extensions of the~$k$-cKP hierarchy and the corresponding solutions were investigated
in~\cite{JMP}.
The latter hierarchies cover matrix generalizations of the Davey--Stewartson (DS) and Nizhnik--Novikov--Veselov (NNV)
systems, $(2+1)$-dimensional extensions of the Yajima--Oikawa and modif\/ied Korteweg--de Vries equations.

Hamiltonian analysis for the above mentioned hierarchies, which is based on group-theoretical and Lie-algebraic methods,
was elaborated in~\cite{BPS, GHP2007,GPP,YaP,CondMat,PBB2007}.
Analytical scheme of the Hamiltonian analysis was described in~\cite{VSG}.

The main aim of this work is to present new $(2+1)$-dimensional extensions of~$k$-cKP and modif\/ied~$k$-cKP hierarchies.
It is organized as follows.
In Section~\ref{extended} we consider reductions~\eqref{ex2+1r1} of Lax operators $L_k$ and $M_n$ involving nonzero
integral terms with degenerate kernels.
The latter reductions allow us to obtain a~hierarchy that is more general then $(2+1)$-dimensional extensions of
the~$k$-cKP hierarchy that we considered in~\cite{JMP}.
This is shown in Remark~\ref{RedSubcase} (Section~\ref{extended}), which describes important special cases of the
obtained hierarchy.
KP hierarchy as a~special case is also included.
In Section~\ref{extended} we also list some nonlinear integrable systems that are provided by Lax pairs~\eqref{ex2+1r1}.
In particular, we get new generalizations of the~$N$-wave problem, the matrix Davey--Stewartson (DS-III) equation and
the matrix KP equation with self-consistent sources (KPSCS).
Despite the fact that the latter systems have more compact scalar counterparts, we also present their matrix versions
due to the recent interest in matrix and, more generally, noncommutative integrable systems (see, e.g.,~\cite{DMH1,DMH2,TodHam,Lecht,Sakhnovich94,Sakhn03,Schiebold1}).

In Section~\ref{section3} we present a~solution generating technique (dressing method) for
hierarchies~\eqref{ex2+1r1}, \eqref{fre1Int} using DTs and BDTs.
New $(2+1)$-dimensional extensions of the modif\/ied~$k$-cKP hierarchies and corresponding solution generating methods are
discussed in Section~\ref{section4}.
Some matrix integrable equations from the respective hierarchy are also listed.
This includes new extensions of the matrix Chen--Lee--Liu equation and the modif\/ied KP equation with self-consistent
sources.
A~short summary of the obtained results and some problems for future investigation are presented in Conclusions.

\section[New $(2+1)$-dimensional generalizations of the~$k$-constrained KP hierarchy]{New $\boldsymbol{(2+1)}$-dimensional generalizations\\
of the~$\boldsymbol{k}$-constrained KP hierarchy}\label{extended}

Further we will use the calculus of the integro-dif\/ferential (pseudo-dif\/ferential) operators of the form
$L=\sum\limits_{i=-\infty}^lf_iD^i$, $l\in{\mathbb{Z}}$ (see, e.g.,~\cite{LDA}).
Coef\/f\/icients $f_i$, $i\in{\mathbb{Z}}$, are matrix-valued functions and symbol ${D}:=\frac{\partial}{\partial x}$
denotes the derivative with respect to~$x$.
Composition (multiplication) of integro-dif\/ferential operators is generated by the commutation rule:
\begin{gather}
{D}^{n}f:= \sum\limits_{j = 0}^\infty \binom{n}{j}  f^{(j)} {D}^{n-j},
\qquad
f^{(j)}:=\frac{\partial^{j}f}{\partial x^j},
\qquad
n\in{\mathbb{Z}},
\label{CommRule}
\end{gather}
where $ \binom{n}{0}: =1$, $\binom{n}{j}:= \frac{n(n - 1)\cdots (n - j + 1)}{j!}$.
${D}^{n}f$ stands for the composition of the operator ${D}^{n}$ and the operator of multiplication by matrix-valued
function~$f$, whereas by curly brackets we will denote the action of the operator on the function, i.e.,
${D}^{n}\{f\}:=f^{(n)}=\frac{\partial^{n}f}{\partial x^n}$.
More generally, we will use notations $L\{f\}$ and $Lf$ in the same manner.

Consider Sato--Zakharov--Shabat dressing operator:
\begin{gather*}
W=I+w_1D^{-1}+w_2D^{-2}+\cdots
\end{gather*}
with $(N\times N)$-matrix-valued coef\/f\/icients $w_i$.
Introduce two dif\/ferential operators $\beta_k\partial_{\tau_k}-\mathcal{J}_kD^k$ and
$\alpha_n\partial_{t_n}-{\tilde{\mathcal{J}}}_nD^n$, $\alpha_n,\beta_k\in{\mathbb{C}}$, $n,k\in{\mathbb{N}}$, where
$\mathcal{J}_k$ and ${\tilde{\mathcal{J}}}_n$ are $N\times N$ commuting matrices (i.e.,
$[{\tilde{\mathcal{J}}}_n,\mathcal{J}_k]=0$).
It is evident that dressed operators have the form:
\begin{gather*}
 L_k:= W\big(\beta_k\partial_{\tau_k}- \mathcal{J}_kD^k\big)W^{-1}
=\beta_k\partial_{\tau_k}-B_k-u_{-1}D^{-1}-u_{-2}D^{-2}-\cdots,
\\
M_n:=W\big(\alpha_n\partial_{t_n}-{\tilde{\mathcal{J}}}_nD^n\big)W^{-1}=\alpha_n\partial_{t_n}-{A}_n-v_{-1}D^{-1}-v_{-2}D^{-2}-\cdots,
\\
B_k=\sum\limits_{j=0}^{k}u_jD^j, \qquad u_k=\mathcal{J}_k,
\qquad
{A}_n=\sum\limits_{i=0}^{n}v_iD^i, \qquad v_n={\tilde{\mathcal{J}}}_n,
\end{gather*}
where $u_j$ and $v_i$ are matrix-valued functions of dimension $N\times N$.
Impose the following reduction on the integral parts of operators $L_k$ and $M_n$:
\begin{gather}
\label{ex2+1r1}
 L_k=\beta_k\partial_{\tau_k}-B_k-{\bf q}{\mathcal M}_0D^{-1}{\bf r}^{\top},
\qquad
M_{n}=\alpha_n\partial_{t_n}-{A}_n-{\tilde{\bf q}}{\tilde{\mathcal M}}_0D^{-1}{\tilde{{\bf r}}}^{\top},
\end{gather}
where ${\bf q}$ and ${\bf r}$ are matrix-valued functions with dimension $N\times m$; ${\tilde{\bf q}}$ and ${\tilde{\bf
r}}$ are matrix-valued functions with dimension $N\times{\tilde{m}}$.
${\mathcal M}_0$ and ${\tilde{\mathcal M}}_0$ are constant matrices with dimensions $m\times m$ and
$\tilde{m}\times{\tilde{m}}$ respectively.

Reductions~\eqref{ex2+1r1} generalize the corresponding analogues obtained in~\cite{JMP}.
It will be shown at the end of this section (reductions~\eqref{Reduction} and Remark~\ref{RedSubcase}).
Moreover, Lax pairs given by~$L_k$ and~$M_n$~\eqref{ex2+1r1} remain covariant
under the action of Darboux and Binary Darboux Transformations (see Section~\ref{section3}),
which allows to construct families of solutions for the corresponding integrable systems.
Lax pairs~\eqref{ex2+1r1} can be also considered as matrix $(2+1)$-dimensional gene\-ra\-li\-zations of the respective operators
appearing in the study of dispersive analogues of Benny's equations~\cite{EOR}.

For technical purposes we will use the following statement:
\begin{Proposition}
\label{TechProp}
For matrix-valued functions $h_1$ and $h_2$ and differential operator $A=\sum\limits_{i=0}^lf_iD^i$, $l\in{\mathbb{N}}$,
with matrix-valued coefficients $f_i$ the following formulae hold:
\begin{gather}
   Ah_1{D}^{-1} h_2=\big(Ah_1{D}^{-1} h_2\big)_{\geq0}+ A\{h_1\}{D}^{-1} h_2,
\label{f1}
\\
   h_1{D}^{-1} h_2 A=\big(h_1{D}^{-1} h_2 A\big)_{\geq0}+ h_1{D}^{-1} \big[A^\tau \big\{h_2^{\top}\big\}\big]^\top,
\label{f2}
\\
   {D}^{-1} h_1 h_2{D}^{-1}=D^{-1}\{h_1h_2\}{D}^{-1} -{D}^{-1}D^{-1}\{h_1h_2\}.
\label{f3}
\end{gather}
\end{Proposition}
Symbol $^{\top}$ in the latter stands for the matrix transpose and $A^{\tau}$ denotes the transpose of~$A$, i.e.,
$A^{\tau}:=\sum\limits_{i=0}^l(-1)^iD^if_i^{\top}$.
Subscript $_{\geq0}$ denotes the dif\/ferential part of the respective operator (e.g.,
$\big(\sum\limits_{i=-\infty}^lf_iD^i\big)_{\geq0}=\sum\limits_{i=0}^lf_iD^i$).
\begin{proof}
All three formulae are consequences of the commutation rule~\eqref{CommRule}.
Let us check~\eqref{f2} and~\eqref{f3}.
It is enough to prove~\eqref{f2} for $A=f_lD^l$.
Using~\eqref{CommRule} we get
\begin{gather*}
h_1{D}^{-1} h_2 A-h_1{D}^{-1} \big[A^\tau \big\{h_2^{\top}\big\}\big]^\top=h_1{D}^{-1} h_2 f_lD^l- h_1{D}^{-1}(-1)^n(h_2f_l)^{(l)}
\\
\qquad
  =h_1\sum\limits_{i=0}^{\infty}(-1)^i(h_2f_l)^{(i)}D^{l-i-1}-h_1\sum\limits_{j=0}^{\infty}(-1)^{j+l}(h_2u_l)^{(j+l)}D^{-1-j}
\\
\qquad
  =h_1\sum\limits_{i=0}^{l-1}(-1)^i(h_2f_l)^{(i)}D^{l-i-1}=\big(h_1{D}^{-1} h_2 A\big)_{\geq0}.
\end{gather*}
\eqref{f3} follows from the following computations:
\begin{gather*}
{D}^{-1} h_1 h_2{D}^{-1}+{D}^{-1}D^{-1}\{h_1h_2\}=\sum\limits_{i=0}^{\infty}(-1)^i(h_1h_2)^{(i)}D^{-i-2}
\\
\qquad
  +\sum\limits_{j=0}^{\infty}(-1)^j(h_1h_2)^{(j-1)}D^{-j-1}=(h_1h_2)^{(-1)}{D}^{-1}=D^{-1}\{h_1h_2\}{D}^{-1}.  \tag*{\qed}
\end{gather*}
\renewcommand{\qed}{}
\end{proof}

\begin{Remark}
Formulae~\eqref{f1}--\eqref{f3} can be found in~\cite{JMP, Syd2004}.
Since the respective references do not contain the proof, we decided to present it in order to make the paper
self-contained.

\end{Remark}
The following statement follows from Proposition~\ref{TechProp}.
\begin{Proposition}
\label{prop-GEN}
Assume that the following equations hold:
\begin{gather}
\label{fre1Int}
L_k\{{\tilde{\bf q}}\}={\tilde{\bf q}}{\Lambda}_{{\tilde{\bf q}}},
\qquad
L_k^{\tau}\{\tilde{\bf r}\}={\tilde{\bf r}}\Lambda_{{\tilde{\bf r}}},
\qquad
M_n\{{\bf q}\}={{\bf q}}{\Lambda}_{{{\bf q}}},
\qquad
M^{\tau}_n\{{\bf r}\}={{\bf r}}{\Lambda}_{{{\bf r}}},
\end{gather}
where ${\Lambda}_{\bf q}$, ${\Lambda}_{\bf r}$ and ${\Lambda}_{\tilde{\bf q}}$, ${\Lambda}_{\tilde{\bf r}}$ are constant
matrices with dimensions $(m\times m)$ and $(\tilde{m}\times\tilde{m})$ respectively, that satisfy equations:
$\Lambda_{\tilde{\bf q}}\tilde{{\cal M}}_0-\tilde{{\cal M}}_0\Lambda^{\top}_{\tilde{\bf r}}=0$, $\Lambda_{\bf q}{{\cal
M}}_0-{{\cal M}}_0\Lambda^{\top}_{\bf r}=0$.

Then Lax equation $[L_k,M_{n}]=0$ holds if and only if equation $[L_k,M_{n}]_{\geq0}=0$ is satisfied.
\end{Proposition}

\begin{proof}
From the equality $[L_k,M_{n}]=[L_k,M_{n}]_{\geq0}+[L_k,M_{n}]_{<0}$ we obtain that Lax equation $[L_k,M_{n}]=0$ is
equivalent to the following one:
\begin{gather*}
[L_k,M_{n}]_{\geq0}=0,\qquad [L_k,M_{n}]_{<0}=0.
\end{gather*}
Thus, it is suf\/f\/icient to prove that equalities $L_k\{{\tilde{\bf q}}\}={\tilde{\bf q}}{\Lambda}_{{\tilde{\bf q}}}$,
$L_k^{\tau}\{\tilde{\bf r}\}={\tilde{\bf r}}\Lambda_{{\tilde{\bf r}}}$, $M_n\{{\bf q}\}={{\bf q}}{\Lambda}_{{{\bf q}}}$,
$M^{\tau}_n\{{\bf r}\}={{\bf r}}{\Lambda}_{{{\bf r}}}$ imply $[L_k,M_{n}]_{<0}=0$.
From the form of operators $L_k$, $M_n$~\eqref{ex2+1r1} we obtain:
\begin{gather}
[L_k,M_{n}]_{<0}=\big[\tilde{{\bf q}}{\tilde{{\mathcal M}}}_0D^{-1}\tilde{{\bf r}}^{\top},\beta_k\partial_{\tau_k}-B_k\big]_{<0}
\nonumber
\\
\phantom{[L_k,M_{n}]_{<0}=}
+\big[{\bf q}{\mathcal M}_0D^{-1}{\bf r}^{\top},{\tilde{{\bf q}}}\tilde{{\mathcal M}}_0D^{-1}{\tilde{{\bf
r}}}^{\top}\big]_{<0}+\big[\alpha_n\partial_{t_n}-A_n,{\bf q}{\mathcal M}_0D^{-1}{\bf r}^{\top}\big]_{<0}.
\label{Frt1}
\end{gather}

After direct computations for each of the three items at the right-hand side of formula~\eqref{Frt1} we get:
\begin{gather}
\big[{\tilde{{\bf q}}}{\tilde{{\mathcal M}}}_0D^{-1}{\tilde{{\bf r}}}^{\top},\beta_k\partial_{\tau_k}-B_k\big]_{<0}
=-\big(\beta_k{\tilde{{\bf q}}}_{\tau_k}-B_k\{{\tilde{{\bf q}}}\}\big){\tilde{\mathcal M}}_0D^{-1}\tilde{{\bf r}}^{\top}
\nonumber
\\
\qquad{}
-{\tilde{{\bf q}}}{\tilde{\mathcal M}}_0D^{-1}\big(\beta_k{\tilde{{\bf r}}}^{\top}_{\tau_k}+\big(B_k^{\tau}\{\tilde{{\bf r}}\}\big)^{\top}\big),
\nonumber
\\
\big[\alpha_n\partial_{t_n}-A_n,{{{\bf q}}}{\mathcal M}_0D^{-1}{{{\bf r}}}^{\top}\big]_{<0}
=(\alpha_n{{{\bf q}}}_{t_n}-A_n\{{{{\bf q}}}\}){\mathcal M}_0D^{-1}{{\bf r}}^{\top}
\nonumber
\\
\qquad{}
+{{{\bf q}}}{\mathcal M}_0D^{-1}\big(\alpha_n{{{\bf r}}}^{\top}_{t_n}+(A_n^{\tau}\{{{\bf r}}\})^{\top}\big),
\nonumber
\\
\big[{\bf q}{\mathcal M}_0D^{-1}{\bf r}^{\top},{\tilde{{\bf q}}}\tilde{{\mathcal M}}_0D^{-1}{\tilde{{\bf
r}}}^{\top}\big]_{<0}={\bf q}{\mathcal M}_0D^{-1}\big\{{\bf r}^{\top}{\tilde{\bf q}}\big\}\tilde{{\mathcal M}}_0D^{-1}{\tilde{\bf r}}^{\top}
-{\bf q}{\mathcal M}_0D^{-1}D^{-1}\big\{{\bf r}^{\top}{\tilde{\bf q}}\big\}{\tilde{{\mathcal M}}}_0{\tilde{\bf r}}^{\top}\nonumber
\\
\qquad{}
-
{\tilde{\bf q}}{\tilde{\mathcal M}}_0D^{-1}\big\{{\tilde{\bf r}}^{\top}{{\bf q}}\big\}{{\mathcal M}}_0D^{-1}{{\bf r}}^{\top}
+ {\tilde{\bf q}}{\tilde{\mathcal M}}_0D^{-1}D^{-1}\big\{{\tilde{\bf r}}^{\top}{{\bf q}}\big\}{{{\mathcal M}}}_0{{\bf
r}}^{\top}.\label{item21}
\end{gather}
The latter formulae are consequences of~\eqref{f1}--\eqref{f3}.
From formulae~\eqref{Frt1}, \eqref{item21} using~\eqref{fre1Int} we get
\begin{gather*}
[L_k,M_{n}]_{<0}=M_n\{{\bf q}\}{\mathcal M}_0D^{-1}{\bf r}^{\top}-{\bf q}{\mathcal M}_0D^{-1}(M_n^{\tau}\{{\bf
r}\})^{\top}-L_k\{{\tilde{\bf q}}\}{\tilde{{\mathcal M}}}_0D^{-1}{\tilde{\bf r}}^{\top}
\\
\phantom{[L_k,M_{n}]_{<0}=}{}
+{\tilde{\bf q}}{\tilde{{\mathcal M}}}_0D^{-1}(L^{\tau}_k\{{\tilde{\bf r}}\})^{\top}={\bf q}\big(\Lambda_{\bf q}{\mathcal
M}_0-{\mathcal M}_0\Lambda_{\bf r}^{\top}\big)D^{-1}{\bf r}^{\top}
\\
\phantom{[L_k,M_{n}]_{<0}=}{}
-{{\tilde{\bf q}}}\big(\Lambda_{\tilde {\bf q}}{\tilde{{\mathcal M}}}_0-{\tilde{{\mathcal M}}}_0\Lambda_{\tilde {\bf
r}}^{\top}\big)D^{-1}{\tilde{\bf r}}^{\top}=0.
\end{gather*}
From the last formula we obtain that equality $[L_k,M_{n}]=0$ is equivalent to condition~\eqref{fre1Int}.
\end{proof}

Consider some nonlinear systems that hierarchy given by~\eqref{ex2+1r1} and~\eqref{fre1Int} contains.
In all examples listed below we assume that equations~\eqref{fre1Int} hold.
Due to Proposition~\ref{prop-GEN} it implies the equivalence of equations $[L_k,M_{n}]=0$ and $[L_k,M_{n}]_{\geq0}=0$.
For simplicity we set ${\Lambda}_{\bf q}={\Lambda}_{\bf r}=0$, ${\Lambda}_{\tilde{\bf q}}={\Lambda}_{\tilde{\bf r}}=0$.

1. $k=1$, $n=1$.
We shall use the following notation $\beta:=\beta_1$, $\alpha:=\alpha_1$, $\tau:=\tau_1$, $t:=t_1$.
Then~\eqref{ex2+1r1} reads:
\begin{gather*}
{{L}_{1}}=\beta\partial_{\tau}-JD+[J,Q]-{\bf{q}}{{\mathcal{M}}_{0}}{{D}^{-1}}{{\bf{r}}^{\top}},
\qquad
{{M}}_{1}=\alpha {\partial}_{{{t}}}-\tilde{J}D+[\tilde{J},Q]-{\tilde{\bf q}}{\tilde{\cal {M}}}_0{{D}^{-1}}{\tilde{\bf
r}}^{\top},
\end{gather*}
where matrices~$J$ and ${\tilde{J}}$ commute.
According to Proposition~\ref{prop-GEN} the commutator equation $[L_1,M_{1}]=0$ is equivalent to the system:
\begin{gather*}
\beta [{\tilde{J}},Q_{\tau}]-\alpha[J,Q_{t}]+JQ_x\tilde{J}-\tilde{J}Q_xJ+[[J,Q],[{\tilde{J}},Q]]
+\big[J,{\tilde{\bf q}}{\tilde{\cal M}}_0{\tilde{\bf r}}^{\top}\big]+\big[{{\bf q}}{{\cal M}}_0{{\bf r}}^{\top},\tilde{J}\big]=0,
\\
\beta{\tilde{\bf q}}_{\tau}-J{\tilde{\bf q}}_x+[J,Q]{\tilde{\bf q}}-{\bf q}{\cal M}_0S_1=0,
\qquad
-\beta{\tilde{\bf r}}^{\top}_{\tau}+{\tilde{\bf r}}^{\top}_xJ+{\tilde{\bf r}}^{\top}[J,Q]+S_2{{{\cal M}}}_0{{\bf
r}}^{\top}= 0,
\\
\alpha{{\bf q}}_{t}-{\tilde J}{\bf q}_x+[{\tilde{J}},Q]{\bf q}-{\tilde{\bf q}}{\tilde{\cal M}}_0S_2=0,
\qquad
-\alpha{{\bf r}}^{\top}_{t}+{\bf r}^{\top}_x{\tilde{J}}+{\bf r}^{\top}[{\tilde{J}},Q]+S_1{\tilde{{{\cal
M}}}}_0{\tilde{{\bf r}}}^{\top}= 0,
\\
S_{1,x}={\bf r}^{\top}{\tilde{\bf q}},
\qquad
S_{2,x}={\tilde{\bf r}}^{\top}{{\bf q}}.
\end{gather*}
The latter system is a~generalization of the~$N$-wave problem~\cite{AblHab,ZakhSh,ZakhShab, ZakhManakN}.
In case we set $Q=0$ we obtain a~noncommutative generalization of the nonlinear system of four
waves~\cite{Ismailov2010,Ismailov2011}.
Under the Hermitian conjugation reduction ${\tilde{\bf r}}={\bar {\tilde{\bf q}}}$, ${\cal M}_0={\cal M}_0^*$,
${\tilde{\cal M}}_0={\tilde{\cal M}}_0^*$, ${{\bf r}}={\bar {\bf q}}$, $Q=-Q^*$, $\alpha,\beta\in{\mathbb{R}}$, $J=J^*$,
${\tilde{J}}={\tilde{J}}^*$ the latter system reads:
\begin{gather*}
 \beta [{\tilde{J}},Q_{\tau}]-\alpha[J,Q_{t}]+JQ_x\tilde{J}-\tilde{J}Q_xJ+[[J,Q],[{\tilde{J}},Q]]
\\
\qquad{}
+ [J,{\tilde{\bf q}}{\tilde{\cal M}}_0{\tilde{\bf q}}^{*}]- [{\tilde{\bf q}}{\tilde{\cal M}}_0{\tilde{\bf
q}}^{*},\tilde{J}]=0,
\qquad
S_{1,x}={\bf q}^{*}{\tilde{\bf q}},
\\
\beta{\tilde{\bf q}}_{\tau}-J{\tilde{\bf q}}_x+[J,Q]{\tilde{\bf q}}-{\bf q}{\cal M}_0S_1=0,
\qquad
\alpha{{\bf q}}_{t}-{\tilde J}{\bf q}_x+[{\tilde{J}},Q]{\bf q}-{\tilde{\bf q}}{\tilde{\cal M}}_0S_1^*=0.
\end{gather*}

2.~$k=1$, $n=2$.
\begin{gather*}
{{L}_{1}}=\beta_1\partial_{\tau_1}-{\bf{q}}{{\mathcal{M}}_{0}}{{D}^{-1}}{{\bf{r}}^{\top}},
\qquad
{{M}}_{2}=\alpha_2 {\partial}_{{{t_2}}}-cD^2+v-{\tilde{\bf q}}{\tilde{\cal {M}}}_0{{D}^{-1}}{\tilde{\bf r}}^{\top},
\qquad
c\in {\mathbb{C}}.
\end{gather*}
Lax equation $[L_1,M_{2}]=0$ is equivalent to the following generalization of the DS-III equation:
\begin{gather*}
 \beta_1{\tilde{\bf q}}_{\tau_1}={\bf q}{\cal M}_0S_1,
\qquad
\beta_1{\tilde{\bf r}}^{\top}_{\tau_1}=S_2{{{\cal M}}}_0{{\bf r}}^{\top},
\qquad
S_{1,x}={\bf r}^{\top}{\tilde{\bf q}},
\\
\alpha_2{{\bf q}}_{t_2}-c{\bf q}_{xx}+v{\bf q}={\tilde{\bf q}}{\tilde{\cal M}}_0S_2,
\qquad
\alpha_2{{\bf r}}^{\top}_{t_2}+c{\bf r}^{\top}_{xx}-{\bf r}^{\top}v=S_1{\tilde{{{\cal M}}}}_0{\tilde{{\bf r}}}^{\top},
\qquad
S_{2,x}={\tilde{\bf r}}^{\top}{\bf q},
\\
\beta_1 v_{\tau_1}=2\big({\bf q}{\cal M}_0{\bf r}^{\top}\big)_x.
\end{gather*}
If we set ${\tilde{\bf q}}=0$, ${\tilde{\bf r}}=0$ we recover the system
\begin{gather*}
 \alpha_2{{\bf q}}_{t_2}-c{\bf q}_{xx}+v{\bf q}=0,
\qquad
\alpha_2{{\bf r}}^{\top}_{t_2}+c{\bf r}^{\top}_{xx}-{\bf r}^{\top}v=0,
\qquad
\beta_1 v_{\tau_1}=2\big({\bf q}{\cal M}_0{\bf r}^{\top}\big)_x,
\end{gather*}
which under reduction $\alpha_2\in\mathrm{i}{\mathbb{R}}$, $\beta_1,c\in{\mathbb{R}}$, ${\cal M}_0={\cal M}_0^*$,
$v=v^*$, ${\bf r}={\bar {\bf q}}$ becomes the matrix version of the DS-III system (see~\cite{Fokas}):
\begin{gather*}
\alpha_2{{\bf q}}_{t_2}-c{\bf q}_{xx}+v{\bf q}=0,
\qquad
\beta_1 v_{\tau_1}=2\big({\bf q}{\cal M}_0{\bf q}^{*}\big)_x.
\end{gather*}

3.~$k=3$, $n=2$.
In this case we obtain the following pair of operators:
\begin{gather*}
{{L}_{3}}=\beta_3\partial_{\tau_3}-c_1\big(D^3-wD-u\big)-{\bf{q}}{{\mathcal{M}}_{0}}{{D}^{-1}}{{\bf{r}}^{\top}},
\\
{{M}}_{2}=\alpha_2 {\partial}_{{{t_2}}}-c_2\big(D^2-v\big)-{\tilde{\bf q}}{\tilde{\cal {M}}}_0{{D}^{-1}}{\tilde{\bf r}}^{\top}.
\end{gather*}
Equation $[L_3,M_2]=0$ is equivalent to the following system:
\begin{gather}
c_1c_2(2w-3v)=0,
\qquad
-\alpha_2c_1w_{t_2}-\frac32c_1c_2v_{xx}+3c_1\big({\tilde{\bf q}}{\tilde{\cal M}}_0{\tilde{\bf r}}^{\top}\big)_x +2c_1c_2u_x=0,
\nonumber
\\
\beta_3c_2v_{\tau_3}-c_1c_2v_{xxx}+3c_1\big({\tilde{\bf q}}_{x}{\tilde{\cal M}}_0{\tilde{\bf r}}^{\top}\big)_x+c_1c_2wv_x-
c_1\big[w,{\tilde{\bf q}}{\tilde{\cal M}}_0{\tilde{\bf r}}^{\top}\big]
\nonumber
\\
\qquad{}
+c_1c_2[u,v]+c_1c_2u_{xx} -\alpha_2 c_1u_{t_2}-2c_2\big({{\bf q}}{{\cal M}}_0{{\bf r}}^{\top}\big)_x=0,
\nonumber
\\
\beta_3{\tilde{{\bf q}}}_{\tau_3}-c_1{\tilde{\bf q}}_{xxx}+c_1w{\tilde{\bf q}}_x+c_1u{\tilde{\bf q}}- {{\bf q}}{{\cal
M}}_0S_1=0,
\qquad
S_{1,x}={{\bf r}}^{\top}{\tilde{\bf q}},
\nonumber
\\
\qquad{}
-\beta_3{\tilde{{\bf r}}}^{\top}_{\tau_3}+c_1{\tilde{\bf r}}^{\top}_{xxx}-c_1\big({\tilde{\bf r}}^{\top}w\big)_x
+c_1{\tilde{\bf r}}^{\top}u+ S_2{{\cal M}}_0{{\bf r}}^{\top}=0,
\qquad
S_{2,x}={\tilde{{\bf r}}}^{\top}{\bf q},
\nonumber
\\
\alpha_2{\bf q}_{t_2}-c_2{\bf q}_{xx}+c_2v{\bf q}-{\tilde{\bf q}}{\tilde{\cal M}}_0S_2=0,
\nonumber
\\
\alpha_2{\bf r}^{\top}_{t_2}+c_2{\bf r}^{\top}_{xx}-c_2{\bf r}^{\top}v- S_1{\tilde{\cal M}}_0{\tilde{\bf r}}^{\top}=0.
\label{eqMain}
\end{gather}
The latter consists of several special cases:

a) $c_1=c_2=1$.
In this case the latter system can be rewritten in the following way:
\begin{gather*}
 -\frac32\alpha_2v_{t_2}-\frac32v_{xx}+3\big({\tilde{\bf q}}{\tilde{\cal M}}_0{\tilde{\bf r}}^{\top}\big)_x +2u_x=0,
\\
\left(\beta_3v_{\tau_3}-\frac14v_{xxx}+\frac32vv_x\right)_x -3\alpha^2v_{t_2t_2}
+\big([u,v]-\big[w,{\tilde{\bf q}}{\tilde{\cal M}}_0{\tilde{\bf r}}^{\top}\big]\big)_x
\\
\qquad{}
+\frac32\big({\tilde{\bf q}}_{xx}{\tilde{\cal M}}_0{\tilde{\bf r}}^{\top}- {\tilde{\bf q}}{\tilde{\cal M}}_0{\tilde{\bf
r}}^{\top}_{xx}+{\alpha} \big({\tilde{\bf q}}{\tilde{\cal M}}_0{\tilde{\bf r}}^{\top}\big)_{t_2}\big)_x-2\big({\bf q}{\cal M}_0{\bf r}^{\top}\big)_{xx} =0,
\\
\beta_3{\tilde{{\bf q}}}_{\tau_3}-{\tilde{\bf q}}_{xxx}+\frac32v{\tilde{\bf q}}_x+u{\tilde{\bf q}}- {{\bf q}}{{\cal M}}_0S_1=0,
\qquad
S_{1,x}={{\bf r}}^{\top}{\tilde{\bf q}},
\\
-\beta_3{\tilde{{\bf r}}}^{\top}_{\tau_3}+{\tilde{\bf r}}^{\top}_{xxx}-\frac32\big({\tilde{\bf r}}^{\top}v\big)_x
+{\tilde{\bf r}}^{\top}u+ S_2{{\cal M}}_0{{\bf r}}^{\top}=0,
\qquad
S_{2,x}={\tilde{{\bf r}}}^{\top}{\bf q},
\\
\alpha_2{\bf q}_{t_2}-{\bf q}_{xx}+v{\bf q}-{\tilde{\bf q}}{\tilde{\cal M}}_0S_2= 0,
\qquad
\alpha_2{\bf r}^{\top}_{t_2}+{\bf r}^{\top}_{xx}-{\bf r}^{\top}v- S_1{\tilde{\cal M}}_0{\tilde{\bf r}}^{\top}=0.
\end{gather*}
In the scalar case ($N=1$) under the Hermitian conjugation reduction: $\alpha_2\in i{\mathbb{R}}$, ${\bf r}={\bar{\bf
q}}$, ${\cal M}_0={\cal M}_0^*$ ($M_2=M_2^*$) and $\beta_3\in{\mathbb{R}}$, ${\tilde{\cal M}}_0=-{\tilde{\cal M}}^*_0$,
$\tilde{{\bf r}}={\bar{\tilde{\bf q}}}$, $w=w^*$, $w^*_x=u+u^*$, $v=v^*$ ($L_3=-L_3^*$).
the latter equation reads:
\begin{gather}
\left(\beta_3v_{\tau_3}-\frac14v_{xxx}+\frac32vv_x\right)_x -3\alpha^2v_{t_2t_2}
\nonumber
\\
\qquad{}
+\frac32\big({\tilde{\bf q}}_{xx}{\tilde{\cal M}}_0{\tilde{\bf q}}^{*}- {\tilde{\bf q}}{\tilde{\cal M}}_0{\tilde{\bf
q}}^{*}_{xx}+{\alpha} \big({\tilde{\bf q}}{\tilde{\cal M}}_0{\tilde{\bf q}}^{*}\big)_t\big)_x-2\big({\bf q}{\cal M}_0{\bf q}^{*}\big)_{xx} =0,
\nonumber
\\
\beta_3{\tilde{{\bf q}}}_{\tau_3}-{\tilde{\bf q}}_{xxx}+\frac32v{\tilde{\bf q}}_x+u{\tilde{\bf q}}- {{\bf q}}{{\cal
M}}_0S_1=0,
\qquad
S_{1,x}={{\bf q}}^{*}{\tilde{\bf q}},
\nonumber
\\
\alpha_2{\bf q}_{t_2}-{\bf q}_{xx}+v{\bf q}-{\tilde{\bf q}}{\tilde{\cal M}}_0S_1^*= 0.
\label{KPSCS_n}
\end{gather}
This system is a~generalization of the KP equation with self-consistent sources (KPSCS).
In particular, if we set ${\tilde{\cal M}}_0=0$, ${\tilde{\bf q}}=0$ we recover KPSCS of the f\/irst type
\begin{gather*}
\left(\beta_3v_{\tau_3}-\frac14v_{xxx}+\frac32vv_x\right)_x -3\alpha^2v_{t_2t_2}=2\big({\bf q}{\cal M}_0{\bf q}^{*}\big)_{xx},
\qquad
\alpha_2{\bf q}_{t_2}-{\bf q}_{xx}+v{\bf q}=0.
\end{gather*}
In case ${{\cal M}}_0=0$, ${\bf q}=0$ in~\eqref{KPSCS_n} we obtain KPSCS of the second type
\begin{gather*}
\left(\beta_3v_{\tau_3}-\frac14v_{xxx}+\frac32vv_x\right)_x -3\alpha^2v_{t_2t_2}
= -\frac32\big({\tilde{\bf q}}_{xx}{\tilde{\cal M}}_0{\tilde{\bf q}}^{*}- {\tilde{\bf q}}{\tilde{\cal M}}_0{\tilde{\bf
q}}^{*}_{xx}+{\alpha} \big({\tilde{\bf q}}{\tilde{\cal M}}_0{\tilde{\bf q}}^{*}\big)_t\big)_x,
\\
\beta_3{\tilde{{\bf q}}}_{\tau_3}-{\tilde{\bf q}}_{xxx}+\frac32v{\tilde{\bf q}}_x+u{\tilde{\bf q}}=0.
\end{gather*}

KPSCS and the respective matrix $(1+1)$-dimensional counterpart (KdV equation with self-consistent sources) have been
investigated recently via Darboux transformations~\cite{Rflows,Rflows-2} and the inverse scattering
method~\cite{Bondarenko}.

b) $c_1=0$, $c_2=1$.
In this case~\eqref{eqMain} becomes the following:
\begin{gather*}
 \beta_3v_{\tau_3}=2\big({{\bf q}}{{\cal M}}_0{{\bf r}}^{\top}\big)_x,
\qquad
\beta_3{\tilde{{\bf q}}}_{\tau_3}- {{\bf q}}{{\cal M}}_0S_1=0,
\qquad
-\beta_3{\tilde{{\bf r}}}^{\top}_{\tau_3}+S_2{{\cal M}}_0{{\bf r}}^{\top}= 0,
\\
\alpha_2{\bf q}_{t_2}-{\bf q}_{xx}+v{\bf q}-{\tilde{\bf q}}{\tilde{\cal M}}_0S_1= 0,
\qquad
S_{1,x}={{\bf r}}^{\top}{\tilde{\bf q}},
\\
\alpha_2{\bf r}^{\top}_{t_2}+{\bf r}^{\top}_{xx}-{\bf r}^{\top}v- S_2{\tilde{\cal M}}_0{\tilde{\bf r}}^{\top}=0,
\qquad
S_2={\tilde{\bf r}}^{\top}{{\bf q}}.
\end{gather*}
In case $\alpha_2\in{\textrm{i}}{\mathbb{R}}$, $\beta_3\in{\mathbb{R}}$, ${\cal M}_0={\cal M}_0^*$, ${\bf q}={\bar{\bf
r}}$, ${\tilde{\cal M}}_0=0$, ${\tilde{\bf q}}={\tilde{\bf r}}=0$ the latter becomes the noncommutative generalization
of the DS-III system.

Now we will show that $(2+1)$-BD$k$-cKP hierarchy presented in~\cite{JMP} can be recovered from Lax
operators~\eqref{ex2+1r1}.
At f\/irst, let us put in formulae~\eqref{ex2+1r1}:
\begin{gather}
{\tilde{\bf q}}:=({\tilde{\bf q}}_1,c_l{\bf q}[0],c_l{\bf q}[1],\ldots,c_l{\bf q}[l]),
\qquad
{\tilde{\bf r}}:=({\tilde{\bf r}}_1,{\bf r}[l],{\bf r}[l-1],\ldots,{\bf r}[0]),
\nonumber
\\
{\tilde{\cal {M}}}_0={\textrm{diag}}({\tilde{\cal {M}}}_1,I_{l+1}\otimes{\cal{M}}_0),
\label{Reduction}
\end{gather}
where ${\bf q}[j]=(L_k)^j\{{\bf q}\}$, ${\bf r}[j]=(L_k^{\tau})^j\{{\bf r}\}$, $j=\overline{0,l}$.
I.e., ${\tilde{m}}=\tilde{m}_1+m(l+1)$ and matrices~${\tilde{\bf q}}$ and~${\tilde{\bf r}}$ consist of
$N\times{\tilde{m}}_1$-matrix-valued blocks~${\tilde{\bf q}}_1$ and~${\tilde{\bf r}}_1$ and $(N\times m)$-matrix-valued
blocks~${\bf q}[j]$ and~${\bf r}[j]$, $j=\overline{0,l}$.
${\tilde{\cal {M}}}_0$ is a~block-diagonal matrix and $I_{l+1}\otimes{\cal{M}}_0$ stands for the tensor product of the
$(l+1)$-dimensional identity matrix $I_{l+1}$ and matrix ${\cal{M}}_0$.
Then we get the following operators in~\eqref{ex2+1r1}:
\begin{gather}
L_k=\beta_k\partial_{\tau_k}-B_k-{\bf q}{\mathcal M}_0D^{-1}{\bf r}^{\top},
\qquad
B_k=\sum\limits_{j=0}^{k}u_jD^j,u_j=u_j(x,\tau_k,t_n),
\qquad
\beta_k\in{\mathbb{C}},
\nonumber
\\
M_n=M_{n,l}=\alpha_n\partial_{t_n}-{A}_n-{\tilde{{\bf q}}}_1{\tilde{\mathcal M}}_1D^{-1}{\tilde{\bf
r}}_1^{\top}-c_l\sum\limits_{j=0}^l{\bf q}[j]{\mathcal M}_0D^{-1}{\bf r}^{\top}[l-j],
\qquad
l=1,\ldots
\nonumber
\\
{A}_n=\sum\limits_{i=0}^{n}v_iD^i, v_i=v_i(x,\tau_k,t_n),
\qquad
\alpha_n\in{\mathbb{C}}.
\label{ex2+1}
\end{gather}
The following proposition holds:
\begin{Proposition}
Assume that equations
\begin{gather}
M_{n,l}\{{\bf q}\}=c_l(L_k)^{l+1}\{{\bf q}\},
\qquad
M_{n,l}^{\tau}\{{\bf{r}}\}=c_l(L_k^{\tau})^{l+1}\{{\bf{r}}\},
\nonumber
\\
L_k\{{\tilde{\bf q}}_1\}={\tilde{\bf q}}_1{\Lambda}_{{\tilde{\bf q}}_1},
\qquad
L_k^{\tau}\{\tilde{\bf r}_1\}={\tilde{\bf r}}_1\Lambda_{{\tilde{\bf r}}_1}
\label{L_eq}
\end{gather}
with constant matrices ${\Lambda}_{{\tilde{\bf q}}_1}$ and $\Lambda_{{\tilde{\bf r}}_1}$ are satisfied, where the latter
solve $\Lambda_{\tilde{\bf q}_1}\tilde{{\cal M}}_1-\tilde{{\cal M}}_1\Lambda^{\top}_{\tilde{\bf r}_1}=0$.
Then Lax equation $[L_k,M_{n,l}]=0$ holds if and only if its differential part is equal to zero, i.e.,
$[L_k,M_{n,l}]_{\geq0}=0$.
\end{Proposition}
\begin{proof}
The proof is similar to the proof of the Proposition~\ref{prop-GEN} and the proof of the Theorem~1 in~\cite{JMP}.
\end{proof}

\begin{Remark}
\label{RedSubcase}
Setting ${\tilde{\bf q}}_1=0$ and ${\tilde{\bf r}}_1=0$ in~\eqref{ex2+1} we recover $(2+1)$-BD$k$-cKP hierarchy (Lax
pairs~\eqref{ex2+1} with equations~\eqref{L_eq}) that contains the following subcases:
\begin{enumerate}\itemsep=0pt
\item $\beta_k=0$, $c_l=0$.
Under this assumption we obtain matrix~$k$-constrained KP hierarchy~\cite{Oevel93}.
We shall point out that the case $\beta_k=0$ and $c_l\neq0$ also leads to matrix~$k$-constrained KP hierarchy.
\item $c_l=0$, $N=1$, $v_n=u_k=1$, $v_{n-1}=u_{k-1}=0$.
In this way we recover $(2+1)$-dimensional~$k$-cKP hierarchy~\cite{6SSS}.
\item $n=0$.
The dif\/ferential part of ${M}_{0,l}$~\eqref{ex2+1} is equal to zero in this case ($A_0=0$) and we get a~new
generalization of DS-III hierarchy.
\item $c_l=0$.
We obtain $(t_A,\tau_B)$-matrix KP hierarchy that was investigated in~\cite{Zeng}.
\item If $l=0$ we recover $(\gamma_A,\sigma_B)$-matrix KP hierarchy~\cite{YYQB}.
\item Case $N=1$, ${\bf q}=0$, ${\bf r}=0$ leads to KP hierarchy.
\end{enumerate}
\end{Remark}

\section[Dressing methods for the new $(2+1)$-dimensional generalizations of~$k$-constrained KP hierarchy]{Dressing methods
for the new $\boldsymbol{(2+1)}$-dimensional\\ generalizations of~$\boldsymbol{k}$-constrained KP hierarchy}
\label{section3}

\subsection{Dressing via Darboux transformations}

In this section we will consider Darboux transformations (DT) for the pair of operators~\eqref{ex2+1r1} and its
reduction~\eqref{ex2+1}.
At f\/irst, we shall start with the linear problem associated with the operator $L_k$~\eqref{ex2+1r1}:
\begin{gather*}
L_k\{\varphi_1\}=\beta_k(\varphi_1)_{\tau_k}-\sum\limits_{j=0}^{k}u_j(\varphi_1)^{(j)}-{\bf q}{\mathcal M}_0D^{-1}\big\{{\bf
r}^{\top}\varphi_1\big\}=\varphi_1\Lambda_1,
\end{gather*}
where $\varphi_1$ is $(N\times N)$-matrix-valued function; $\Lambda_1$ is a~constant matrix with dimension $N\times N$.
Introduce the DT in the following way:
\begin{gather}
\label{W}
W_1[\varphi_1]=\varphi_1 D\varphi_1^{-1}=D-\varphi_{1,x}\varphi_{1}^{-1}.
\end{gather}
The following proposition holds.
\begin{Proposition}
\label{D1}
The operator $\hat{L}_{k}[1]:=W_1[\varphi_1]L_kW_1^{-1}[\varphi_1]$ obtained from $L_k$~\eqref{ex2+1r1} via DT~\eqref{W}
has the form
\begin{gather*}
\hat{L}_{k}[1]:=W_1[\varphi_1]L_kW_1^{-1}[\varphi_1]=\beta_k\partial_{\tau_k}-\hat{B}_k-\hat{\bf q}_1{\mathcal M}_0D^{-1}{\hat{{\bf r}}}_1^{\top},
\qquad
\hat{B}_k[1]=\sum\limits_{j=0}^{k}\hat{u}_{j}[1]D^j,
\end{gather*}
where
\begin{gather*}
\hat{{\bf q}}_1=W_1[\varphi_1]\{{\bf q}\},
\qquad
{\hat{\bf r}}_1=W_1^{-1,\tau}[\varphi_1]\{{\bf r}\}.
\end{gather*}
$\hat{u}_j[1]$ are $(N\times N)$-matrix coefficients depending on function $\varphi_1$ and coefficients $u_i$,
$i=\overline{0,k}$.
In particular, $\hat{u}_{k}[1]=u_k$.
\end{Proposition}
\begin{proof}
It is evident that the inverse operator to~\eqref{W} has the form $W_1^{-1}[\varphi_1]=\varphi_1D^{-1}\varphi_1^{-1}$.
Thus, we have
\begin{gather*}
\hat{L}_{k}[1]=W_1[\varphi_1]L_kW_1^{-1}[\varphi_1]=\varphi_1D\varphi_1^{-1}\big(\beta_k\partial_{\tau_k}-{B}_k-{\bf
q}{\cal M}_0D^{-1}{{{\bf r}}}^{\top}\big)\varphi_1D^{-1}\varphi_1^{-1}
\\
\phantom{\hat{L}_{k}[1]}
=\beta_k\partial_{\tau_k}+(\hat{L}_{k}[1])_{\geq0}+(\hat{L}_{k}[1])_{<0},
\end{gather*}
where $(\hat{L}_{k}[1])_{\geq0}=-\hat{B}_{k}[1]=-\sum\limits_{j=0}^{k}\hat{u}_{j}[1]D^j$. It remains to f\/ind the
explicit form of $(\hat{L}_k)_{<0}$.
Using formulae~\eqref{f1}--\eqref{f3} we have:
\begin{gather*}
 (\hat{L}_{k}[1])_{<0}=\big(\beta_k\varphi_1D\varphi_1^{-1}\varphi_{1,\tau_k}D^{-1}\varphi_1^{-1}\big)_{<0}-
\big(\varphi_1D\varphi_1^{-1}B_k\{\varphi_1\}D^{-1}\varphi_1^{-1}\big)_{<0}
\\
\phantom{(\hat{L}_{k}[1])_{<0}=}{}
-\big(\varphi_1D\varphi_1^{-1}{\bf q}{\mathcal M}_0D^{-1}\big\{{\bf
r}^{\top}\varphi_1\big\}D^{-1}\varphi_1^{-1}-\varphi_1D\varphi_1^{-1}{\bf q}{\mathcal M}_0D^{-1}D^{-1}\big\{{\bf
r}^{\top}\varphi_1\big\}\varphi_1^{-1}\big)_{<0}
\\
\phantom{(\hat{L}_{k}[1])_{<0}}
=\big(\varphi_1D\varphi_1^{-1}\varphi_1\Lambda_1 D^{-1}\varphi_1^{-1}\big)_{<0}+\varphi_1D\big\{\varphi_1^{-1}{\bf q}\big\}{\mathcal
M}_0D^{-1}D^{-1}\big\{{\bf r}^{\top}\varphi_1\big\}\varphi_1^{-1}
\\
\phantom{(\hat{L}_{k}[1])_{<0}}
=-W_1\{{\bf q}\}{\mathcal M}_0D^{-1}\big(W_1^{-1,\tau}\{{\bf r}\}\big)^{\top}.  \tag*{\qed}
\end{gather*}
\renewcommand{\qed}{}
\end{proof}

It is also possible to generalize the latter theorem to the case of f\/inite number of solutions of linear problems
associated with the operator $L_k$.
Namely, let functions $\varphi_s$, $s=\overline{1,K}$ be solutions of the problems:
\begin{gather}
\label{26}
L_k\{\varphi_s\}=\beta_k(\varphi_s)_{\tau_k}-\sum\limits_{j=0}^{k}u_j(\varphi_s)^{(j)}-{\bf q}{\mathcal M}_0D^{-1}\big\{{\bf
r}^{\top}\varphi_s\big\}=\varphi_s\Lambda_s,
\qquad
s=\overline{1,K}.
\end{gather}
For further convenience we shall use the notations $\varphi_s[1]:=\varphi_s$, $s=\overline{1,K}$ and def\/ine the
following functions:
\begin{gather}
\label{Crum1}
\varphi_s[2]=W_1[\varphi_{1}[1]]\{\varphi_s[1]\},
\qquad
s=\overline{2,K}.
\end{gather}
Now, using functions $\varphi_1[1]$, $\varphi_2[2]$, we shall def\/ine functions $\varphi_s[3]$, $s=\overline{3,K}$:
\begin{gather*}
\varphi_s[3]:=W_1[\varphi_2[2]]\{\varphi_s[2]\}=W_1[\varphi_2[2]]W_1[\varphi_1[1]]\{\varphi_s[1]\},
\qquad
s=\overline{3,K}.
\end{gather*}
At the~$p$-th step we obtain functions: $\varphi_s[p]:=W_1[\varphi_{p-1}[p-1]]\{\varphi_s[p-1]\}
=W_1[\varphi_{p-1}[p-1]]\cdots W_1[\varphi_2[2]]W_1[\varphi_1[1]]\{\varphi_s[1]\}$, $s=\overline{p,K}$. Now we shall
construct the following generalization of DT~\eqref{W}:
\begin{gather}
 W_K[\varphi_1,\ldots,\varphi_K]=W_1[\varphi_K[K]]\cdots W_1[\varphi_1[1]]
\nonumber
\\
\qquad
=\big(D-\varphi_{K,x}[K]\varphi^{-1}_{K}[K]\big) \cdots\big(D-\varphi_{1,x}[1]\varphi_{1}^{-1}[1]\big).
\label{WC}
\end{gather}
The following statement holds:
\begin{Proposition}
\label{Prop}
The operator
\begin{gather*}
\hat{L}_{k}[K]:=W_K[\varphi_1[1],\varphi_2[1],\ldots,\varphi_K[1]]L_kW_K^{-1}[\varphi_1[1],\varphi_2[1],\ldots,\varphi_K[1]]=W_KL_kW_K^{-1}
\end{gather*}
obtained from $L_k$~\eqref{ex2+1} via DT~\eqref{WC} has the form
\begin{gather*}
\hat{L}_{k}[K]:=W_KL_kW_K^{-1}=\beta_k\partial_{\tau_k}-\hat{B}_{k}[K]-\hat{\bf q}_K{\mathcal M}_0D^{-1}{\hat{{\bf
r}}}_K^{\top}, \qquad \hat{B}_{k}[K]=\sum\limits_{j=0}^{k}\hat{u}_{j}[K]D^j,
\end{gather*}
where
\begin{gather*}
 \hat{{\bf q}}_K=W_K[\varphi_1[1],\ldots,\varphi_K[1]]\{{\bf q}\}, \qquad {\hat{\bf
r}}_K=W_K^{-1,\tau}[\varphi_1[1],\ldots,\varphi_K[1]]\{{\bf r}\}.
\end{gather*}
$\hat{u}_{j}[K]$ are $(N\times N)$-matrix coefficients depending on functions $\varphi_s$, $s=\overline{1,K}$ and
coefficients $u_i$, $i=\overline{0,k}$.
In particular, $\hat{u}_{k}[K]=u_k$.
\end{Proposition}
\begin{proof}
The proof can be done via induction by~$K$.
Namely, assume that the statement holds for $K-1$.
I.e.,
\begin{gather}
\label{Lk}
 \hat{L}_{k}[K-1]=W_{K-1}L_kW_{K-1}^{-1}=\beta_k\partial_{\tau_k}-\hat{B}_{k}[K-1]-\hat{\bf q}_{K-1}{\mathcal
M}_0D^{-1}{\hat{{\bf r}}}_{K-1}^{\top},
\end{gather}
with $\hat{{\bf q}}_{K-1}=W_{K-1}[\varphi_1[1],\ldots,\varphi_{K-1}[1]]\{{\bf q}\}$ and ${\hat{\bf
r}}_{K-1}=W_{K-1}^{-1,\tau}[\varphi_1[1],\ldots,\varphi_{K-1}[1]]\{{\bf r}\}$.
The function $\varphi_{K}[K]=W_{K-1}\{\varphi_K[1]\}=W_{K-1}[\varphi_1,\ldots,\varphi_{K-1}]\{\varphi_{K}[1]\}$ (see
formulae~\eqref{Crum1},~\eqref{WC}) satis\-f\/ies the equation:
$\hat{L}_{k}[K-1]\{\varphi_{K}[K]\}=W_{K-1}L_kW_{K-1}^{-1}\{W_{K-1}\{\varphi_K[1]\}\}=$
\newline
$=W_{K-1}L_k\{\varphi_{K}[1]\}=\varphi_K[K]\Lambda_K$.

Now, it remains to apply  Proposition~\ref{D1} to operator $\hat{L}_{k}[K-1]$~\eqref{Lk} with the DT
$W_1[\varphi_K[K]]$ (see formula~\eqref{W}) and use formula $W_K=W_{1}[\varphi_K[K]]W_{K-1}$ that immediately follows
from~\eqref{WC}.
\end{proof}
\begin{Remark}
We shall also point out that in a~scalar case ($N=1$) the DT $W_K$~\eqref{WC} can be rewritten in the following way:
\begin{gather}
\label{DetForm}
W_K:=\frac{1}{{\mathcal{W}}[\varphi_1,\varphi_2,\ldots,\varphi_K]}\left|
\begin{array}{cccc}
\varphi_1     &\dots& \varphi_K & 1
\\
\varphi'_1     &\dots& \varphi'_K & {D}
\\
\dots&\dots&\dots&\dots
\\
\varphi_1^{(K)} &\dots& \varphi_K^{(K)} & {D}^K
\end{array}
\right|=D^K+\sum\limits_{i=0}^{K-1}w_iD^i,
\end{gather}
where ${\mathcal{W}}[\varphi_1,\varphi_2,\ldots,\varphi_K]$ denotes the Wronskian constructed by solutions $\varphi_j$,
$j=1,\ldots,K$, of the linear problem~\eqref{26}.
It acts on the vector-valued function ${\bf q}=(q_1,\ldots,q_m)$ in the following way: $W_K\{{\bf
q}\}=(W_K\{q_1\},\ldots,W_K\{q_m\})$, where
$W_K\{q_j\}=\frac{{\mathcal{W}}[\varphi_1,\varphi_2,\ldots,\varphi_K,q_j]}{{\mathcal{W}}[\varphi_1,\varphi_2,\ldots,\varphi_K]}$.
\end{Remark}

The Darboux transformations~\eqref{W},~\eqref{WC} are widely used for the solution generating technique involving Lax
pairs consisting of dif\/ferential operators~\cite{GM,GN, Matveev79,Matveev,OrlovRau}.
Corresponding extensions to integro-dif\/ferential cases of Lax pairs were made in~\cite{LZL2,MSS, Oevel93,Oevel96,6SSS} to
construct solutions of constrained KP hierarchies and their generalizations.
Transformations that generalize~\eqref{W},~\eqref{WC} also arise in the bidif\/ferential calculus approach to integrable
systems and their hierarchies~\cite{FMH1,FMH2}.
In contrast to~\eqref{WC}, formula~\eqref{DetForm} does not require iterative applications of DTs and therefore can be
used more ef\/fectively in the scalar case.
We point out that in the matrix (noncommutative) case~\eqref{DetForm} is not valid anymore.
However, the corresponding quasideterminant representations can be used (see, e.g.,~\cite{GM, GN}).

From Proposition~\ref{Prop} we obtain the corollary for Lax pairs~\eqref{ex2+1r1}.
Namely, let functions $\varphi_s$, $s=\overline{1,K}$ be solutions of the problems:
\begin{gather*}
 L_k\{\varphi_s\}=\beta_k(\varphi_s)_{\tau_k}-\sum\limits_{j=0}^{k}u_j(\varphi_s)^{(j)}-{\bf q}{\mathcal
M}_0D^{-1}\big\{{\bf r}^{\top}\varphi_s\big\}=\varphi_s\Lambda_s,
\\
M_n\{\varphi_s\}=\alpha_n(\varphi_s)_{t_n}-\sum\limits_{i=0}^{n}v_i(\varphi_s)^{(i)}-{\tilde{\bf q}}{\tilde{\mathcal
M}}_0D^{-1}\big\{{\tilde{{\bf r}}}^{\top}\varphi_s\big\}=\varphi_s{\tilde{\Lambda}}_s,
\qquad
s=\overline{1,K}.
\end{gather*}

Then the following statement holds:
\begin{Corollary}
Assume that Lax equation with operators $L_k$ and $M_n$~\eqref{ex2+1r1} holds: $[L_k,M_n]=0$.
Then:
\begin{enumerate}\itemsep=0pt
\item[$1.$] Transformed operators
\begin{gather*}
\hat{L}_{k}[K]:=W_K[\varphi_1,\ldots,\varphi_K]L_kW^{-1}_K[\varphi_1,\ldots,\varphi_K],\\
\hat{M}_{n}[K]:=W_K[\varphi_1,\ldots,\varphi_K]M_nW^{-1}_K[\varphi_1,\ldots,\varphi_K],
\end{gather*}
 where $W_K$ is defined
by~\eqref{WC}, have the form:
\begin{gather}
\hat{L}_{k}[K]:=W_KL_kW_K^{-1}=\beta_k\partial_{\tau_k}-\hat{B}_{k}[K]-\hat{\bf q}_K{\mathcal M}_0D^{-1}{\hat{{\bf r}}}_K^{\top},
\nonumber
\\
\hat{M}_{n}[K]:=W_KM_nW_K^{-1}=\alpha_n\partial_{t_n}-\hat{A}_{n}[K]-\hat{{\tilde{{\bf q}}}}_K{\tilde{\mathcal
M}}_0D^{-1}{\hat{\tilde{{\bf r}}}}_K^{\top},
\nonumber
\\
\hat{A}_{n}[K]=\sum\limits_{i=0}^{n}\hat{v}_{i}[K]D^i, \qquad \hat{B}_{k}[K]=\sum\limits_{j=0}^{k}\hat{u}_{j}[K]D^j,
\label{TKP}
\end{gather}
where
\begin{gather*}
 \hat{{\bf q}}_K=W_K[\varphi_1[1],\ldots,\varphi_K[1]]\{{\bf q}\}, \qquad {\hat{\bf
r}}_K=W_K^{-1,\tau}[\varphi_1[1],\ldots,\varphi_K[1]]\{{\bf r}\},
\\
\hat{{\tilde{{\bf q}}}}_K=W_K[\varphi_1[1],\ldots,\varphi_K[1]]\{{\tilde{{\bf q}}}\}, \qquad {\hat{\tilde{\bf
r}}}_K=W_K^{-1,\tau}[\varphi_1[1],\ldots,\varphi_K[1]]\{{\tilde{\bf r}}\}.
\end{gather*}
\item[$2.$] The operators $\hat{L}_{k}[K]$ and $\hat{M}_{n}[K]$~\eqref{TKP} satisfy Lax equation:
$[\hat{L}_{k}[K],\hat{M}_{n}[K]]=0$.
\item[$3.$] In case of reduction~\eqref{Reduction} in Lax pair~\eqref{ex2+1r1} we have:
\begin{gather*}
\hat{{\tilde{{\bf
q}}}}_K=({\hat{\tilde{\bf q}}}_{1,K},c_l{\hat{\bf q}}_K[0],\ldots,c_l\hat{\bf q}_K[l]),\qquad
\hat{{\tilde{{\bf
r}}}}_K=({\hat{\tilde{\bf r}}}_{1,K},\hat{\bf r}_K[0],\ldots\hat{\bf r}_K[l]),
\end{gather*}
 where
 \begin{gather*}
 {\hat{\tilde{\bf
q}}}_{1,K}=W_K\{{\tilde{\bf q}}_1\} , \qquad
{\hat{\tilde{\bf r}}}_{1,K}=W^{-1,\tau}_K\{{\tilde{\bf r}}_1\} ,
\\
\hat{{\bf q}}[j]=(\hat{L}_{k}[K])^j\{\hat{{\bf q}}_K\} , \qquad \hat{{\bf r}}[j]=(\hat{L}_{k}^{\tau}[K])^j\{\hat{{\bf
r}}_K\} , \qquad j=\overline{0,l}.
\end{gather*}
\end{enumerate}
\end{Corollary}
\begin{proof}
1.~Form~\eqref{TKP} of operators $\hat{L}_{k}[K]$, $\hat{M}_{n}[K]$ follows from Proposition~\ref{Prop}.

2.~We obtain the proof of this item from the following formulae
\begin{gather*}
\big[\hat{L}_{k}[K],\hat{M}_{n}[K]\big]=\big[W_KL_kW_K^{-1},W_KM_nW_K^{-1}\big]=W_K[L_k,M_n]W_K^{-1}=0 .
\end{gather*}

3.~From formulae:
\begin{gather*}
 \hat{\tilde{{\bf q}}}_K=W_K\{({\tilde{\bf q}}_1,c_l{\bf q}[0],\ldots,c_l{\bf q}[l])\}=(W_K\{{\tilde{\bf
q}}_1\},c_lW_K\{{\bf q}[0]\},\ldots, c_lW_K\{{\bf q}[l]\}),
\\
W_K\{{\bf q}[j]\}=W_K\{(L_k)^{j}\{{\bf q}\}\}=\big(W_KL_kW_K^{-1}\big)\{W_K\{{\bf q}\}\}=\hat{L}_{k}[K]\{{\hat{\bf q}}_K\}
\end{gather*}
we get the form of ${\hat{\tilde{\bf q}}}_K$ mentioned in item 3.
The form of ${\hat{\tilde{\bf r}}}_K$ can be obtained in a~similar way.
\end{proof}

\subsection{Dressing via binary Darboux transformations}

In this section we will show that results of paper~\cite{K2009} on binary Darboux transformations (BDT) for linear
integro-dif\/ferential operators can be extended to families of Lax pairs given by~\eqref{ex2+1r1}.
Namely, let $(N\times K)$-matrix functions~$\varphi$ and~$\psi$ be solutions of linear problems:
\begin{gather*}
L_k\{\varphi\}=\varphi\Lambda_k,
\qquad
L_k^{\tau}\{\psi\}=\psi\tilde{\Lambda}_k,
\qquad
\Lambda_k,\tilde{\Lambda}_k\in {\rm Mat}_{K\times K}({\mathbb{C}}).
\end{gather*}

Following~\cite{K2009} we introduce BDT in the following way:
\begin{gather}
\label{Wbin}
W=I-\varphi\big(C+D^{-1}\big\{\psi^{\top}\varphi\big\}\big)^{-1}D^{-1}\psi^{\top},
\end{gather}
where~$C$ is a~$K\times K$-constant nondegenerate matrix.
The inverse operator $W^{-1}$ has the form:
\begin{gather*}
W^{-1}=I+\varphi D^{-1}\big(C+D^{-1}\{\psi^{\top}\varphi\}\big)^{-1}\psi^{\top}.
\end{gather*}
The following theorem is proven in~\cite{K2009}.
\begin{Theorem}\label{2009}
The operator $\hat{L}_k:=WL_kW^{-1}$ obtained from $L_k$ in~\eqref{ex2+1r1} via BDT~\eqref{Wbin} has the form
\begin{gather*}
\hat{L}_k:=WL_kW^{-1}=\beta_k\partial_{\tau_k}-\hat{B}_k-\hat{\bf q}{\cal M}_0D^{-1}{\hat{{\bf r}}}^{\top}+\Phi{\cal M}_kD^{-1}\Psi^{\top},
\qquad
\hat{B}_k=\sum\limits_{j=0}^{k}\hat{u}_jD^j,
\end{gather*}
where
\begin{gather*}
{\cal M}_k=C\Lambda_k-\tilde{\Lambda}^{\top}_kC,
\qquad
\Phi=\varphi\Delta^{-1},
\qquad
\Psi=\psi\Delta^{-1,\top},
\qquad
\Delta=C+D^{-1}\big\{\psi^{\top}\varphi\big\},
\\
{\hat{\bf q}}=W\{{\bf q}\},
\qquad
{\hat{\bf r}}=W^{-1,\tau}\{{\bf r}\}.
\end{gather*}
$\hat{u}_j$ are $(N\times N)$-matrix coefficients depending on functions~$\varphi$,~$\psi$ and $u_j$.
In particular,
\begin{gather*}
\hat{u}_k=u_k,
\qquad
\hat{u}_{k-1}=u_{k-1}+\big[u_k,\varphi\big(C+D^{-1}\big\{\psi^{\top}\varphi\big\}\big)^{-1}\psi^{\top}\big].
\end{gather*}
\end{Theorem}

Solution generating method for the hierarchy~\eqref{ex2+1r1}--\eqref{fre1Int} is given by the corollary, which follows
from the previous theorem.
\begin{Corollary}
\label{CorolMain}
Let $(N\times K)$-matrix functions~$\varphi$ and~$\psi$ satisfy equations:
\begin{gather*}
L_k\{\varphi\}=\varphi\Lambda_{k,1}, L_k^{\tau}\{\psi\}=\psi\tilde{\Lambda}_{k,1},
\qquad
\Lambda_{k,1},\tilde{\Lambda}_{k,1}\in {\rm Mat}_{K\times K}({\mathbb{C}}),
\\
M_n\{\varphi\}=\varphi\Lambda_{n,2}, M_k^{\tau}\{\psi\}=\psi\tilde{\Lambda}_{n,2},
\qquad
\Lambda_{n,2},\tilde{\Lambda}_{n,2}\in {\rm Mat}_{K\times K}({\mathbb{C}})
\end{gather*}
with operators $L_k$ and $M_n$~\eqref{ex2+1r1} satisfying $[L_k,M_n]=0$.
Then transformed operators~$\hat{L}_k$ and~$\hat{M}_n$ satisfy Lax equation $[\hat{L}_k,\hat{M}_n]=0$ and have the form:
\begin{gather}
 \hat{L}_k:=WL_kW^{-1}=\beta_k\partial_{\tau_k}-\hat{B}_k-\hat{\bf q}{\cal M}_0D^{-1}{\hat{{\bf r}}}^{\top}+\Phi{\cal
M}_{k,1}D^{-1}\Psi^{\top}, \qquad \hat{B}_k=\sum\limits_{j=0}^{k}\hat{u}_jD^j,
\nonumber
\\
\hat{M}_{n}:=WM_nW^{-1}=\alpha_n\partial_{t_n}-\hat{A}_{n}-\hat{{\tilde{{\bf q}}}}{\tilde{\mathcal
M}}_0D^{-1}{\hat{\tilde{{\bf r}}}}^{\top}+\Phi{{{\cal M}}}_{n,2}D^{-1}\Psi^{\top},
\qquad\!
\hat{A}_{n}=\sum\limits_{i=0}^{n}\hat{v}_{i}D^i,\!\!\!
\label{LopMop}
\end{gather}
where
\begin{gather*}
{\cal M}_{k,1}=C\Lambda_{k,1}-\tilde{\Lambda}^{\top}_{k,1}C,
\qquad
{{{\cal M}}}_{n,2}=C\Lambda_{n,2}-\tilde{\Lambda}^{\top}_{n,2}C, \qquad \Phi=\varphi\Delta^{-1}, \qquad \Psi=\psi\Delta^{-1,\top},
\\
\Delta=C+D^{-1}\{\psi^{\top}\varphi\}, \qquad {\hat{\bf q}}=W\{{\bf q}\}, \qquad {\hat{\bf r}}=W^{-1,\tau}\{{\bf r}\},
{\hat{\tilde{\bf q}}}=W\{{\tilde{\bf q}}\}, \qquad {\hat{\tilde{\bf r}}}=W^{-1,\tau}\{{\tilde{\bf r}}\}.
\end{gather*}
$\hat{u}_j$ are $(N\times N)$-matrix coefficients depending on functions~$\varphi$,~$\psi$ and $u_j$, $v_i$.
In particular,
\begin{gather*}
 \hat{u}_k=u_k,
\hat{u}_{k-1}=u_{k-1}+\big[u_k,\varphi\big(C+D^{-1}\big\{\psi^{\top}\varphi\big\}\big)^{-1}\psi^{\top}\big],
\\
\hat{v}_n=v_n,
\hat{v}_{n-1}=v_{n-1}+\big[v_n,\varphi\big(C+D^{-1}\big\{\psi^{\top}\varphi\big\}\big)^{-1}\psi^{\top}\big].
\end{gather*}
\end{Corollary}
\begin{proof}
From formulae $W[L_k,M_n]W^{-1}=[\hat{L}_k,\hat{M}_n]=0$ we obtain that Lax equation with transformed operators is
satisf\/ied.
Form~\eqref{LopMop} of the transformed operators $\hat{L}_k$ and $\hat{M}_n$ follows from Theorem~\ref{2009}.
\end{proof}

BDTs were used to generate solutions of the constrained KP hierarchies in~\cite{NJJC, Will97} and they were also applied
to $(2+1)$-BD$k$-cKP hierarchy in~\cite{JMP}.
Theorem~\ref{2009} with Corollary~\ref{CorolMain} extend the respective results to $(2+1)$-dimensional generalizations of
the~$k$-constrained KP hierarchy~\eqref{ex2+1r1},~\eqref{fre1Int}.

\section[New $(2+1)$-dimensional generalizations of the modif\/ied~$k$-constrained KP hierarchy]{New $\boldsymbol{(2+1)}$-dimensional generalizations\\
of the modif\/ied~$\boldsymbol{k}$-constrained KP hierarchy}
\label{section4}

We will investigate the following $(2+1)$-dimensional generalizations of the modif\/ied~$k$-constrained KP ($k$-cmKP)
hierarchy:
\begin{gather}
L_k=\beta_k\partial_{\tau_k}-B_k-{\bf q}{\mathcal M}_0D^{-1}{\bf r}^{\top}D,
\qquad
B_k=\sum\limits_{j=1}^{k}u_jD^j,
\quad
u_j=u_j(x,\tau_k,t_n),
\quad
\beta_k\in{\mathbb{C}},
\nonumber
\\
M_{n}=\alpha_n\partial_{t_n}-{A}_n-{\tilde{\bf q}}{\tilde{\mathcal M}}_0D^{-1}{\tilde{{\bf r}}}^{\top}D,
\qquad
{A}_n=\sum\limits_{i=1}^{n}v_iD^i,
\quad
v_i=v_i(x,\tau_k,t_n), \quad \alpha_n\in{\mathbb{C}},\!\!\!\!
\label{ex2+1r12}
\end{gather}
where $u_j$ and $v_i$ are matrix-valued functions of dimension $N\times N$; ${\bf q}$ and ${\bf r}$ are matrix-valued
functions of dimension $N\times m$; ${\tilde{\bf q}}$ and ${\tilde{\bf r}}$ are matrix-valued functions with dimension
$N\times{\tilde{m}}$.
${\mathcal M}_0$ and ${\tilde{\mathcal M}}_0$ are constant matrices with dimensions $m\times m$ and
$\tilde{m}\times{\tilde{m}}$ respectively.

Using equality~\eqref{f2} we can rewrite operators~\eqref{ex2+1r12} as
\begin{gather}
 L_k=\beta_k\partial_{\tau_k}-B_k-{\bf q}{\mathcal M}_0{\bf r}^{\top}+{\bf q}{\mathcal M}_0D^{-1}{\bf r}^{\top}_x,
\nonumber
\\
M_{n}=\alpha_n\partial_{t_n}-{A}_n-{\tilde{\bf q}}{\tilde{\mathcal M}}_0{\tilde{{\bf r}}}^{\top}
+{\tilde{\bf q}}{\tilde{\mathcal M}}_0D^{-1}{\tilde{{\bf r}}}^{\top}_x.
\label{ex2+1r12r}
\end{gather}
From the latter it becomes clear that~\eqref{ex2+1r12} can be considered as a~reduction in~\eqref{ex2+1r1}
($B_k\rightarrow B_k+{\bf q}{\mathcal M}_0{\bf r}^{\top}$, ${\bf r}^{\top}\rightarrow -{\bf r}^{\top}_x$).
However, Lax pairs~\eqref{ex2+1r12} provide us with dif\/ferent $(2+1)$-dimensional equations (see~\eqref{CLL_gen}
and~\eqref{KPSCS_mod}).
In addition, they require dif\/ferent kind of BDT (see Theorem~\ref{Theor} with its corollary) for the corresponding
solution generating technique.

Setting ${\tilde{{\bf q}}}=0$, ${\tilde{{\bf r}}}=0$, $N=1$ in~\eqref{ex2+1r12} we recover $(2+1)$-dimensional~$k$-cmKP
hierarchy~\cite{LZL2}.

The following proposition holds:
\begin{Proposition}
Lax equation $[L_k,M_{n}]=0$ is satisfied in case the following equations hold:
\begin{gather}
 [L_k,M_{n}]_{\geq0}=0,
\qquad
L_k\{{\tilde{\bf q}}\}={\tilde{\bf q}}{\Lambda}_{{\tilde{\bf q}}},
\qquad
\big(D^{-1}L_k^{\tau}D\big)\{\tilde{\bf r}\}={\tilde{\bf r}}\Lambda_{{\tilde{\bf r}}},
\nonumber
\\
M_n\{{\bf q}\}={{\bf q}}{\Lambda}_{{{\bf q}}},
\qquad
\big(D^{-1}M^{\tau}_nD\big)\{{\bf r}\}={{\bf r}}{\Lambda}_{{{\bf r}}},
\label{fre1mod}
\end{gather}
where ${\Lambda}_{\bf q}$, ${\Lambda}_{\bf r}$, ${\Lambda}_{\tilde{\bf q}}$, ${\Lambda}_{\tilde{\bf r}}$ are constant
matrices with dimensions $(m\times m)$ and $(\tilde{m}\times\tilde{m})$ respectively that satisfy equations:
$\Lambda_{\tilde{\bf q}}\tilde{{\cal M}}_0-\tilde{{\cal M}}_0\Lambda^{\top}_{\tilde{\bf r}}=0$, $\Lambda_{\bf q}{{\cal
M}}_0-{{\cal M}}_0\Lambda^{\top}_{\bf r}=0$.
\end{Proposition}
\begin{proof}
The proof is a~direct consequence of reductions~\eqref{ex2+1r12r} and Proposition~\ref{prop-GEN}.
\end{proof}
Consider some examples of the hierarchy given by~\eqref{ex2+1r12} and~\eqref{fre1mod}.

1.~$k=1$, $n=2$.
\begin{gather*}
 L_1=\beta_1\partial_{\tau_1}-D-{\bf q}{\mathcal M}_0D^{-1}{\bf r}^{\top}D,
\\
M_{2}=\alpha_2\partial_{t_2}-D^2-vD-{\tilde{\bf q}}{\tilde{\mathcal M}}_0D^{-1}{\tilde{{\bf r}}}^{\top}D.
\end{gather*}
Lax representation $[L_1,M_2]=0$ is equivalent to the following system:
\begin{gather*}
 \alpha_2{\bf q}_{t_2}-{\bf q}_{xx}-v{\bf q}_x-{\tilde{\bf q}}{\tilde{\cal M}}_0D^{-1}\big\{{\tilde{\bf r}}^{\top}{\bf
q}_x\big\}={\bf q}\Lambda_{\bf q},
\\
-\alpha_2{\bf r}_{t_2}-{\bf r}_{xx}+v^{\top}{\bf r}_x-{\tilde{\bf r}}{\tilde{\cal M}}_0^{\top}D^{-1}\big\{{\tilde{\bf
q}}^{\top}{\bf r}_x\big\}={\bf r}\Lambda_{\bf r},
\\
\beta_1{\tilde{\bf q}}_{\tau_1}-{\tilde{\bf q}}_x-{\bf q}{\cal M}_0D^{-1}\big\{{\bf r}^{\top}{\tilde{\bf q}}_x\big\}={\tilde{\bf
q}}\Lambda_{\tilde{\bf q}},
\qquad
-\beta_1{\tilde{\bf r}}_{{\tau}_1}+{\tilde{\bf r}}_x-{\bf r}{\cal M}_0^{\top}D^{-1}\big\{{\bf q}^{\top}{\tilde{\bf
r}}_x\big\}={\tilde{\bf r}}\Lambda_{\tilde{\bf r}},
\\
v_x-\beta_1 v_{\tau_1}=2\big({\bf q}{\cal M}_0{\bf r}^{\top}\big)_x.
\end{gather*}
In case of the Hermitian conjugation reduction $\beta_1\in{\mathbb{R}}$, $\alpha_2\in i{\mathbb{R}}$, ${\cal
M}_0^*=-{\cal M}_0$, ${\tilde{\cal M}}_0^*={\tilde{\cal M}}_0$, ${\bar{\tilde{\bf r}}}={\tilde{\bf q}}$, ${\bar{\bf
r}}={\bf q}$, $v=-v^*$ ($L_1^*=-DL_1D^{-1}$, $M_2^*=DM_2D^{-1}$) the latter equation reduces to the following one:
\begin{gather}
 \alpha_2{\bf q}_{t_2}-{\bf q}_{xx}-v{\bf q}_x-{\tilde{\bf q}}{\tilde{\cal M}}_0D^{-1}\{{\tilde{\bf q}}^{*}{\bf
q}_x\}={\bf q}\Lambda_{\bf q},
\nonumber
\\
\beta_1{\tilde{\bf q}}_{\tau_1}-{\tilde{\bf q}}_x-{\bf q}{\cal M}_0D^{-1}\{{\bf q}^{*}{\tilde{\bf q}}_x\}={\tilde{\bf
q}}\Lambda_{\tilde{\bf q}},
\qquad
v_x-\beta_1 v_{\tau_1}=2\big({\bf q}{\cal M}_0{\bf q}^{*}\big)_x.
\label{CLL_gen}
\end{gather}
In case we set ${\tilde{\bf q}}=0$, $\Lambda_{\bf q}=0$ we get a~matrix $(2+1)$-dimensional generalization of the
Chen-Lee-Liu equation
\begin{gather*}
\alpha_2{\bf q}_{t_2}-{\bf q}_{xx}-v{\bf q}_x=0,
\qquad
v_x-\beta_1 v_{\tau_1}=2\big({\bf q}{\cal M}_0{\bf q}^{*}\big)_x.
\end{gather*}

2.~$k=3$, $n=2$.
\begin{gather*}
 L_3=\beta_3\partial_{\tau_3}-c_1\big(D^3+wD^2+vD\big)-{\bf q}{\cal M}_0D^{-1}{\bf r}^{\top}D,
\\
M_2=\alpha_2\partial_{t_2}-c_2D^2-uD-{\tilde{\bf q}}\tilde{{\cal M}_0}D^{-1}{\tilde{\bf r}}^{\top}D.
\end{gather*}
Using~\eqref{fre1mod} we get that Lax representation $[L_3,M_2]=0$ is equivalent to the following system
\begin{gather*}
 3u_x-2c_2w_x-[u,w]=0,
\\
\alpha_2c_1w_{t_2}-c_1c_2w_{xx}+2c_1wu_x-c_1uw_x+3c_1u_{xx}
\\
\qquad{}
+3c_1\big({\tilde{\bf q}}{\tilde{\cal M}}_0{\tilde{\bf r}}^{\top}\big)_x-2c_2c_1v_x-c_2\big[{\tilde{\bf q}}{\tilde{\cal M}}_0{\tilde
{\bf r}}^{\top},w\big]+c_1[v,u]=0,
\\
-\beta_3u_{\tau_3}+c_1u_{xxx}+c_1wu_{xx}+3c_1\big({\tilde{\bf q}}_{x}{\tilde{\cal M}}_0{\tilde{\bf
r}}^{\top}\big)_x+2c_1w{\tilde{\bf q}}_x{\tilde{\cal M}}_0{\tilde{\bf r}}^{\top}
\\
\qquad{}
+c_1{\tilde{\bf q}}{\tilde{\cal M}}_0{\tilde{\bf
r}}^{\top}_xw+\alpha_2c_1v_{t_2}-c_1c_2v_{xx}+c_1vu_x-c_1uv_x+c_1\big[v,{\tilde{\bf q}}{\cal M}_0{\tilde{\bf r}}^{\top}\big]
\\
\qquad{}
-2c_2\big({\bf q}{\cal M}_0{\bf r}^{\top}\big)_x+c_1w{\tilde{\bf q}}{\tilde{\cal M}}_0{\tilde{\bf r}}^{\top}_x+\big[{\bf q}{\cal
M}_0{\bf r}^{\top},u\big]=0,
\\
\beta_3{\tilde{\bf q}}_{\tau_3}-c_1{\tilde{\bf q}}_{xxx}-c_1w{\tilde{\bf q}}_{xx}-c_3v{\tilde{\bf q}}_x-{{\bf q}}{{\cal
M}}_0D^{-1}\big\{{{\bf r}}^{\top}{\tilde{\bf q}}_x\big\}={\tilde{\bf q}}\Lambda_{\tilde{\bf q}},
\\
\alpha_2{\bf q}_{t_2}-c_2{\bf q}_{xx}-u{\bf q}_x-{\tilde{\bf q}}{\tilde{\cal M}}_0D^{-1}\big\{{\tilde{\bf r}}^{\top}{\bf
q}_x\big\}={\bf q}\Lambda_{\bf q},
\\
-\beta_3{\tilde{\bf r}}_{\tau_3}+c_1{\tilde{\bf r}}_{xxx}-c_1\big(w^{\top}{\tilde{\bf r}}_x\big)_x+c_1v^{\top}{\tilde{\bf
r}}_x-{\bf r}{\cal M}_0^{\top}D^{-1}\big\{{\bf q}^{\top}{\tilde{\bf r}}_x\big\}={\tilde{\bf r}}\Lambda_{{\tilde{\bf r}}},
\\
-\alpha_2{\bf r}_{t_2}-c_2{\bf r}_{xx}+u^{\top}{\bf r}_x-{\tilde{\bf r}}{\tilde{\cal M}}_0^{\top}D^{-1}\big\{{\tilde{\bf
q}}^{\top}{\bf r}_x\big\}={\bf r}\Lambda_{\bf r}.
\end{gather*}
Set $c_1=c_2=1$ in the scalar case ($N=1$).
Eliminating variables~$w$ and~$v$ from the f\/irst and second equation respectively, we get
\begin{gather*}
-\beta_3u_{{\tau}_3}-\frac14u_{xxx}-\frac38u^2u_x+\frac34\alpha_2u_xD^{-1}\{u_{t_2}\}+\frac34\alpha_2^2D^{-1}\{u_{t_2t_2}\}
\\
\qquad{}
+\frac32\big(u{\tilde{\bf q}}{\tilde{\cal M}}_0{\tilde{\bf r}}^{\top}\big)_x+\frac32\alpha_2\big({\tilde{\bf q}}{\cal
M}_0{\tilde{\bf r}}^{\top}\big)_{t_2}-\frac32\big({\tilde{\bf q}}{\tilde{\cal M}}_0{\tilde{\bf r}}^{\top}\big)_{xx}-2({\bf q}{\cal
M}_0{\bf r}^{\top})_x=0,
\\
\beta_3{\tilde{\bf q}}_{\tau_3}-{\tilde{\bf q}}_{xxx}-\frac32u{\tilde{\bf q}}_{xx}-v{\tilde{\bf q}}_x-{{\bf q}}{{\cal
M}}_0D^{-1}\big\{{{\bf r}}^{\top}{\tilde{\bf q}}_x\big\}={\tilde{\bf q}}\Lambda_{\tilde{\bf q}},
\\
\alpha_2{\bf q}_{t_2}-{\bf q}_{xx}-u{\bf q}_x-{\tilde{\bf q}}{\tilde{\cal M}}_0D^{-1}\big\{{\tilde{\bf r}}^{\top}{\bf
q}_x\big\}={\bf q}\Lambda_{\bf q},
\\
-\beta_3{\tilde{\bf r}}_{\tau_3}+{\tilde{\bf r}}_{xxx}-\frac32(u{\tilde{\bf r}}_x)_x+v{\tilde{\bf r}}_x-{\bf r}{\cal
M}_0^{\top}D^{-1}\big\{{\bf q}^{\top}{\tilde{\bf r}}_x\big\}={\tilde{\bf r}}\Lambda_{{\tilde{\bf r}}},
\\
-\alpha_2{\bf r}_{t_2}-{\bf r}_{xx}+u{\bf r}_x-{\tilde{\bf r}}{\tilde{\cal M}}_0^{\top}D^{-1}\big\{{\tilde{\bf
q}}^{\top}{\bf r}_x\big\}={\bf r}\Lambda_{\bf r},
\\
v=\frac34u_x+\frac38u^2+\frac32\big({\tilde{\bf q}}{\tilde{\cal M}}_0{\tilde{\bf
r}}^{\top}\big)+\frac34\alpha_2D^{-1}\{u_{t_2}\}.
\end{gather*}
The latter under the Hermitian conjugation reduction $\alpha_2\in i{\mathbb{R}}$, $\beta_3\in{\mathbb{R}}$, ${\cal
M}_0^*=-{\cal M}_0$, ${\tilde{\cal M}}_0={\tilde{\cal M}}_0^*$, $\bar{u}=-u$, ${\bar{\tilde{\bf r}}}={\tilde{\bf q}}$,
${\bar{\bf r}}={\bf q}$ ($L_3^*=-DL_3D^{-1}$, $M_2^*=DM_2D^{-1}$) reads:
\begin{gather}
-\beta_3u_{{\tau}_3}-\frac14u_{xxx}-\frac38u^2u_x+\frac34\alpha_2u_xD^{-1}\{u_{t_2}\}+\frac34\alpha_2^2D^{-1}\{u_{t_2t_2}\}
\nonumber
\\
\qquad
+\frac32(u{\tilde{\bf q}}{\tilde{\cal M}}_0{\tilde{\bf q}}^{*})_x+\frac32\alpha_2({\tilde{\bf q}}{\tilde{\cal M}}_0{\tilde{\bf q}}^{*})_{t_2}
-\frac32({\tilde{\bf q}}{\tilde{\cal M}}_0{\tilde{\bf q}}^{*})_{xx}-2\big({\bf q}{\cal M}_0{\bf q}^{*}\big)_x=0,
\nonumber
\\
\beta_3{\tilde{\bf q}}_{\tau_3}-{\tilde{\bf q}}_{xxx}-\frac32u{\tilde{\bf q}}_{xx}
-\left(\frac34u_x+\frac38u^2+\frac32({\tilde{\bf q}}{\tilde{\cal M}}_0{\tilde{\bf q}}^{*})+\frac34\alpha_2D^{-1}\{u_{t_2}\}\right)
{\tilde{\bf q}}_x
\nonumber
\\
\qquad{}
-{{\bf q}}{{\cal M}}_0D^{-1}\{{{\bf q}}^{*}{\tilde{\bf q}}_x\}={\tilde{\bf q}}\Lambda_{\tilde{\bf q}},
\nonumber
\\
\alpha_2{\bf q}_{t_2}-{\bf q}_{xx}-u{\bf q}_x-{\tilde{\bf q}}{\tilde{\cal M}}_0D^{-1}\{{\tilde{\bf r}}^{\top}{\bf
q}_x\}={\bf q}\Lambda_{\bf q}.
\label{KPSCS_mod}
\end{gather}

${\tilde{\bf q}}=0$, $\Lambda_{{\bf q}}=0$ lead
to the modif\/ied KPSCS of the f\/irst type
\begin{gather*}
-\beta_3u_{{\tau}_3}-\frac14u_{xxx}-\frac38u^2u_x+\frac34\alpha_2u_xD^{-1}\{u_{t_2}\}+\frac34\alpha_2^2D^{-1}\{u_{t_2t_2}\}=2\big({\bf q}{\cal M}_0{\bf q}^{*}\big)_x,
\\
\alpha_2{\bf q}_{t_2}-{\bf q}_{xx}-u{\bf q}_x=0.
\end{gather*}

If ${\bf q}=0$, ${\tilde{\Lambda}}_{{\tilde{\bf q}}}=0$ in~\eqref{KPSCS_mod} we recover the second type of the modif\/ied KPSCS
\begin{gather*}
-\beta_3u_{{\tau}_3}-\frac14u_{xxx}-\frac38u^2u_x+\frac34\alpha_2u_xD^{-1}\{u_{t_2}\}+\frac34\alpha_2^2D^{-1}\{u_{t_2t_2}\}
\\
\qquad{}
+\frac32(u{\tilde{\bf q}}{\tilde{\cal M}}_0{\tilde{\bf q}}^{*})_x+\frac32\alpha_2({\tilde{\bf q}}{\tilde{\cal
M}}_0{\tilde{\bf q}}^{*})_{t_2}-\frac32({\tilde{\bf q}}{\tilde{\cal M}}_0{\tilde{\bf q}}^{*})_{xx}=0,
\\
\beta_3{\tilde{\bf q}}_{\tau_3}-{\tilde{\bf q}}_{xxx}-\frac32u{\tilde{\bf q}}_{xx}
-\left(\frac34u_x+\frac38u^2+\frac32({\tilde{\bf q}}{\tilde{\cal M}}_0{\tilde{\bf q}}^{*})
+\frac34\alpha_2D^{-1}\{u_{t_2}\}\right){\tilde{\bf q}}_x=0,
\end{gather*}
Both types were investigated in~\cite{LZL2} within $(2+1)$-dimensional extensions of the~$k$-cmKP hierarchy.

\subsection{Dressing via binary Darboux transformations}

In this subsection we consider dressing methods for $(2+1)$-dimensional extensions of the modif\/ied~$k$-constrained KP
hierarchy given by~\eqref{ex2+1r12} and~\eqref{fre1mod}.
First of all, we start with the matrix version of the theorem that was proven in~\cite{NonOsc}.
\begin{Theorem}
\label{Theor}
Let $(N\times K)$-matrix functions~$\varphi$ and~$\psi$ satisfy linear problems:
\begin{gather*}
  L_k\{\varphi\}=\varphi\Lambda_k, L_k^{\tau}\{\psi\}=\psi\tilde{\Lambda}_k, \qquad \Lambda_k,\tilde{\Lambda}_k\in
{\rm Mat}_{K\times K}({\mathbb{C}}),
\\
L_k=\beta_k\partial_{\tau_k}-B_k-{\bf q}{\mathcal M}_0D^{-1}{\bf r}^{\top}D,
\qquad
B_k=\sum\limits_{i=1}^ku_iD^i.
\end{gather*}
Then the operator ${L}_k$ transformed via
\begin{gather}
\label{Wm}
W_m:=w_0^{-1}W=w_0^{-1}\big(I-\varphi\Delta^{-1}D^{-1}\psi^{\top}\big)=I-\varphi{\tilde{\Delta}}^{-1}D^{-1}(D^{-1}\{\psi\})^{\top}D,
\end{gather}
where
\begin{gather}
 w_0=I_N-\varphi\Delta^{-1}D^{-1}\big\{\psi^{\top}\big\}, \qquad \tilde{\Delta}=-C+D^{-1}\big\{D^{-1}\big\{\psi^{\top}\big\}\varphi_x\big\},
\nonumber
\\
\Delta=C+D^{-1}\big\{\psi^{\top}\varphi\big\},
\label{addon}
\end{gather}
has the form:
\begin{gather*}
 \hat{L}_k:=W_mL_kW_m^{-1}=\beta_k\partial_{\tau_k}-\hat{B}_k-{\hat{\bf q}}{\mathcal M}_0D^{-1}{\hat{{\bf
r}}}^{\top}D+{{\Phi}}{\mathcal M}_kD^{-1}{{\Psi}}^{\top}D,
\\
\hat{B}_k=\sum\limits_{j=1}^{k}\hat{u}_jD^j, \qquad \hat{u}_k=u_k, \qquad \hat{u}_{k-1}=u_{k-1}+ku_kw^{-1}_0w_{0,x}, \qquad \ldots,
\end{gather*}
where
\begin{gather*}
{\mathcal M}_k=C\Lambda_k-\tilde{\Lambda}_k^{\top}C,
\qquad
{\tilde{\Phi}}=-W_m\{\varphi\}C^{-1}=\varphi {\tilde{\Delta}}^{-1},
\\
{\tilde\Psi}=D^{-1}\big\{W_m^{\tau,-1}\{\psi\}\big\}C^{-1,\top}=D^{-1}\{\psi\}\Delta^{-1,\top}, \qquad {\hat{\bf q}}=W_m\{{\bf q}\},
{\hat{\bf r}}=D^{-1}W_m^{-1,\tau}D\{{\bf r}\}.
\end{gather*}
\end{Theorem}
\begin{proof}
The proof is analogous to the proof of Theorem~2 in~\cite{NonOsc}.
\end{proof}
The following consequence of the latter theorem provides a~solution generating method for the hierarchy given
by~\eqref{ex2+1r12} and~\eqref{fre1mod}:
\begin{Corollary}
Let $(N\times K)$-matrix functions~$\varphi$ and~$\psi$ satisfy linear problems:
\begin{gather*}
L_k\{\varphi\}=\varphi\Lambda_{k,1},
\qquad
L_k^{\tau}\{\psi\}=\psi\tilde{\Lambda}_{k,1},
\qquad
\Lambda_{k,1},
\tilde{\Lambda}_{k,1}\in {\rm Mat}_{K\times K}({\mathbb{C}}),
\\
M_n\{\varphi\}=\varphi\Lambda_{n,2},
\qquad
M_n^{\tau}\{\psi\}=\psi\tilde{\Lambda}_{n,2},
\qquad \Lambda_{n,2},
\tilde{\Lambda}_{n,2}\in {\rm Mat}_{K\times K}({\mathbb{C}})
\end{gather*}
with operators $L_k$ and $M_n$ given by~\eqref{ex2+1r12}.

The operators $\hat{L}_k=W_mL_kW_m^{-1}$ and $\hat{M}_n=W_mM_nW^{-1}_m$ transformed via $W_m$~\eqref{Wm}, \eqref{addon}
have the form:
\begin{gather*}
 \hat{L}_k:=\beta_k\partial_{\tau_k}-\hat{B}_k-{\hat{\bf q}}{\mathcal M}_0D^{-1}{\hat{{\bf
r}}}^{\top}D+{\tilde{\Phi}}{\mathcal M}_{k,1}D^{-1}{\tilde{\Psi}}^{\top}D,
\qquad
\hat{B}_k=\sum\limits_{j=1}^{k}\hat{u}_jD^j,
\\
\hat{M}_{n}=\alpha_n\partial_{t_n}-{\hat{{A}}}_n-{\hat{\bf q}}{\tilde{\mathcal M}}_0D^{-1}{\hat{{\bf
r}}}^{\top}D+{\tilde{\Phi}}{{{\mathcal M}}}_{n,2}D^{-1}{\tilde{\Psi}}^{\top}D,
\qquad
\hat{A}_n=\sum\limits_{i=1}^{n}\hat{v}_iD^i,
\end{gather*}
where
\begin{gather*}
 {\mathcal M}_{k,1}=C\Lambda_{k,1}-\tilde{\Lambda}^{\top}_{k,1}C,
\qquad
{{{\mathcal M}}}_{n,2}=C\Lambda_{n,2}-\tilde{\Lambda}^{\top}_{n,2}C,
\qquad
{\tilde{\Phi}}=-W_m\{\varphi\}C^{-1}=\varphi {\tilde{\Delta}}^{-1},
\\
{\tilde\Psi}=D^{-1}\big\{W_m^{\tau,-1}\{\psi\}\big\}C^{-1,\top}=D^{-1}\{\psi\}\Delta^{-1,\top},
\qquad
{\hat{\bf q}}=W_m\{{\bf q}\},
\\
{\hat{\bf r}}=D^{-1}W_m^{-1,\tau}D\{{\bf r}\},
\qquad
\tilde{\Delta}=-C+D^{-1}\big\{D^{-1}\big\{\psi^{\top}\big\}\varphi_x\big\}.
\end{gather*}
\end{Corollary}

\section{Conclusion}

In this work we proposed new integrable generalizations of the KP and modif\/ied KP hierarchy with self-consistent
sources.
The obtained hierarchies of nonlinear equations include, in particular, matrix integrable system that contains as
special cases two types of the matrix KP equation with self-consistent sources (KPSCS) and its modif\/ied version.
They also cover new generalizations of the~$N$-wave problem and the DS-III system.
Under reductions~\eqref{Reduction} imposed on the obtained hierarchies one recovers $(2+1)$-BD$k$-cKP hierarchy.
The latter contains $(t_A,\tau_B)$- and $(\gamma_A,\sigma_B)$-matrix KP hierarchies~\cite{Zeng,YYQB} (see~\cite{JMP} for
details).

\begin{Remark}
It should be pointed out that in the scalar case ($N=1$) Lax pairs~\eqref{ex2+1r12} admit the following reduction:
\begin{gather*}
{{\bf q}}=\big({{\bf q}}_1,{\bf q}_2,-{\bf q}_2{\cal M}_0{\bf r}^{\top}_2-D^{-1}\{u\},1\big),
\qquad
{{\bf r}}=\big({{\bf r}}_1,D^{-1}\{{{\bf r}}_2\},1,D^{-1}\{u\}\big),
\\
{\tilde{\bf q}}=\big({\tilde{\bf q}}_1,{\tilde{\bf q}}_2,-{\tilde{{\bf q}}}_2{\tilde{\cal M}}_1{\tilde{{\bf
r}}}^{\top}_2-D^{-1}\{{\tilde{u}}\},1,{{\bf q}}_1[0],{{\bf q}}_1[1],\ldots,{{\bf q}}_1[l]\big),
\qquad
{\bf q}_1[j]:=L^j\{{\bf q}_1\},
\\
{\tilde{\bf r}}=\big({\tilde{\bf r}}_1,{\tilde{\bf r}}_2,1,D^{-1}\{{\tilde{u}}\},{{\bf r}}_1[l],{{\bf
r}}_1[l-1],\ldots,{{\bf r}}_1[0]\big),
\qquad
{\bf r}_1[j]:=(L^{\tau})^j\{{\bf r}_1\},
\\
{\cal M}_0=\textrm{diag}({\cal M}_1,{\cal M}_1,1,1),
\qquad
{\tilde{\cal M}}_0=\textrm{diag}({\tilde{\cal M}}_1,{\tilde{\cal M}}_1,1,1, I_{l+1}\otimes {\cal M}_1),
\end{gather*}
where ${\bf q}_j$, ${\bf r}_j$ and ${\tilde{\bf q}}_j$, ${\tilde{\bf r}}_j$, $j=1,2$, are vectors of functions of
dimensions $(1\times m_0)$ and $(1\times {\tilde{m}}_0)$ respectively.
${\cal M}_1$ and ${\tilde{\cal M}}_1$ are square matrices of dimensions $m_0$ and ${\tilde{m}}_0$.
$I_{l+1}\otimes {\cal M}_1$ denotes the tensor product of the identity matrix $I_{l+1}$ and ${\cal M}_1$.

It leads to the following family of integro-dif\/ferential operators in~\eqref{ex2+1r12}
\begin{gather*}
L_k=\beta_k\partial_{\tau_k}-B_k-{\bf q}_1{\mathcal M}_1D^{-1}{\bf r}^{\top}_1D+{\bf q}_2{\mathcal M}_1D^{-1}{\bf
r}^{\top}_2+D^{-1}u,
\\
M_{n}=\alpha_n\partial_{t_n}-{A}_n-{\tilde{\bf q}}_1{\tilde{\mathcal M}}_1D^{-1}{\tilde{{\bf r}}}^{\top}_1D+{\tilde{\bf
q}}_2{\tilde{\mathcal M}}_1D^{-1}{\tilde{{\bf r}}}^{\top}_2+D^{-1}{\tilde{u}}
\\
\qquad{}
-c_l\sum\limits_{j=0}^l{\bf q}_1[j]{\mathcal M}_0D^{-1}{\bf r}^{\top}_1[l-j]D.
\end{gather*}
Lax equation $[L_k,M_n]=0$ involving the latter operators should lead (under additional reductions) to $(2+1)$-dimensional
generalizations of the corresponding integrable systems that were obtained in~\cite{NonOsc}.
In particular it concerns systems that extend KdV, mKdV and Kaup--Broer equations.
\end{Remark}
In this paper we also elaborated solution generating methods for the proposed
hierarchies \eqref{ex2+1r1}, \eqref{fre1Int} and~\eqref{ex2+1r12}, \eqref{fre1mod} respectively via DTs and BDTs.

The latter involve f\/ixed solutions of linear problems and an arbitrary seed (initial) solution of the corresponding
integrable system.
Exact solutions of equations with self-consistent sources (complexitons, negatons, positons) and the underlying
hierarchies were studied in~\cite{Zeng,Ma_Compl,Negaton-positon}.
One of the problems for future interest consists in looking for the corresponding analogues of these solutions in the
obtained generalizations.
The same question concerns lumps and rogue wave solutions that were investigated in several integrable systems
recently~\cite{Rogue1,Rogue2,Rogue3,DF11,DF13,DF12,Rogue4,Rogue5}

It is also known that inverse scattering and spectral methods~\cite{MAblowitz,Calogero,Konopelchenko,NovikovZakharov}
were applied to generate solutions of equations with self-consistent sources~\cite{Bondarenko, Gerdjikov,ScatInv}.
An extension of these methods to the obtained hierarchies and comparison with results that can be provided by BDTs
(e.g., following~\cite{BDTScat}) presents an interest for us.

The search for the corresponding discrete counterparts of the constructed hierarchies is another problem for future
investigation.
The latter is expected to contain the discrete KP equation with self-consistent sources~\cite{DL, HW}.
One of the possible ways to solve the problem consists in looking for the formulation of the corresponding continuous
hierarchy within a~framework of bidif\/ferential calculus.
The latter framework provides better possibilities to search for the discrete counterparts of the corresponding
continuous systems (see, e.g.,~\cite{DMH-NLS}).

\subsection*{Acknowledgements}

The authors are grateful to Professors Folkert M\"uller-Hoissen and Maxim Pavlov for fruitful discussions and useful advice in preparation of this paper.
The authors also wish to express their gratitude to the referees for their valuable comments and suggestions.
O.~Chvartatskyi has been supported via the Alexander von Humboldt foundation.
Yu.M.~Sydorenko is grateful to the Ministry of Education, Science, Youth and Sports of Ukraine for partial f\/inancial
support (Research Grant MA-107F).

\pdfbookmark[1]{References}{ref}
\LastPageEnding

\end{document}